\documentclass[reprint,superscriptaddress,nofootinbib,floatfix]{revtex4-2}

\usepackage{graphicx}
\usepackage{amsmath}
\usepackage{amssymb}

\graphicspath{{figs/}}
\newcommand{\Ohn}{\mathit{Oh}}

\begin{document}

\title{The bursting of hollow droplets}

\author{Alfonso M. Ga\~n\'an-Calvo}
\email{amgc@us.es}
\affiliation{Depto.\ de Ingenier\'{\i}a Aeroespacial y Mec\'anica de Fluidos,
Escuela T\'ecnica Superior de Ingenier\'{\i}a, Universidad de Sevilla,
41092 Sevilla, Spain}

\begin{abstract}
A bubble bursting from a drop of finite size (a hollow droplet) is the
generic bursting event in a breaking wave, yet it has been studied almost
exclusively at unbounded baths. We follow it from the
puncture of the film to the spectrum of a spray, in axisymmetric simulations
over liquid-to-gas volume ratios $\Lambda$ from $1/16$ to $512$ and Ohnesorge
numbers from $0.005$ to $0.11$. Ejection ceases at
$\Ohn_1=\Ohn_c(1+2\beta/\lambda)$, with $\beta\simeq0.83$ and
$\lambda=(1+\Lambda)^{1/3}$: confinement extends ejection to bubbles too small
to eject at a flat surface, and the distance $\delta=1-\Ohn/\Ohn_1$ to that
boundary organizes everything that follows. The collapse, driven by
capillary waves, is that of a gas thread pinching next to its
mouth, whether the liquid shell is punctured once or twice. The ejected volume
is $M_e=C\,\delta\,V_{\rm red}$, with $C\simeq0.013$ and
$V_{\rm red}=V_{\rm gas}V_{\rm liq}/(V_{\rm gas}+V_{\rm liq})$ a reduced
volume. Whatever the confinement, the first droplet is smallest at
$\delta\simeq0.2$ in units of the viscocapillary length
$\ell_\mu=\mu^2/(\rho\sigma)$ and at $\delta\simeq0.3$ in units of the bubble
radius $R_0$, where the unbounded bath also places its optimum. The largest
reaches a tenth of $R_0$. Between them the census is exponential, with
a scale set by $R_0$ and no lower cut-off: the smallest droplet a simulation
records is set by its resolution, not by $\ell_\mu$, which belongs to the pinch
singularity rather than to the droplets. Mixed over bubble populations,
the census gives spray spectra in closed form, whose shape measures
the population. The bursting also produces hollow droplets, each a
potential new generator.
\end{abstract}

\maketitle

\section{Introduction}\label{sec:intro}

\begin{figure*}
  \centering
  \includegraphics[width=\textwidth]{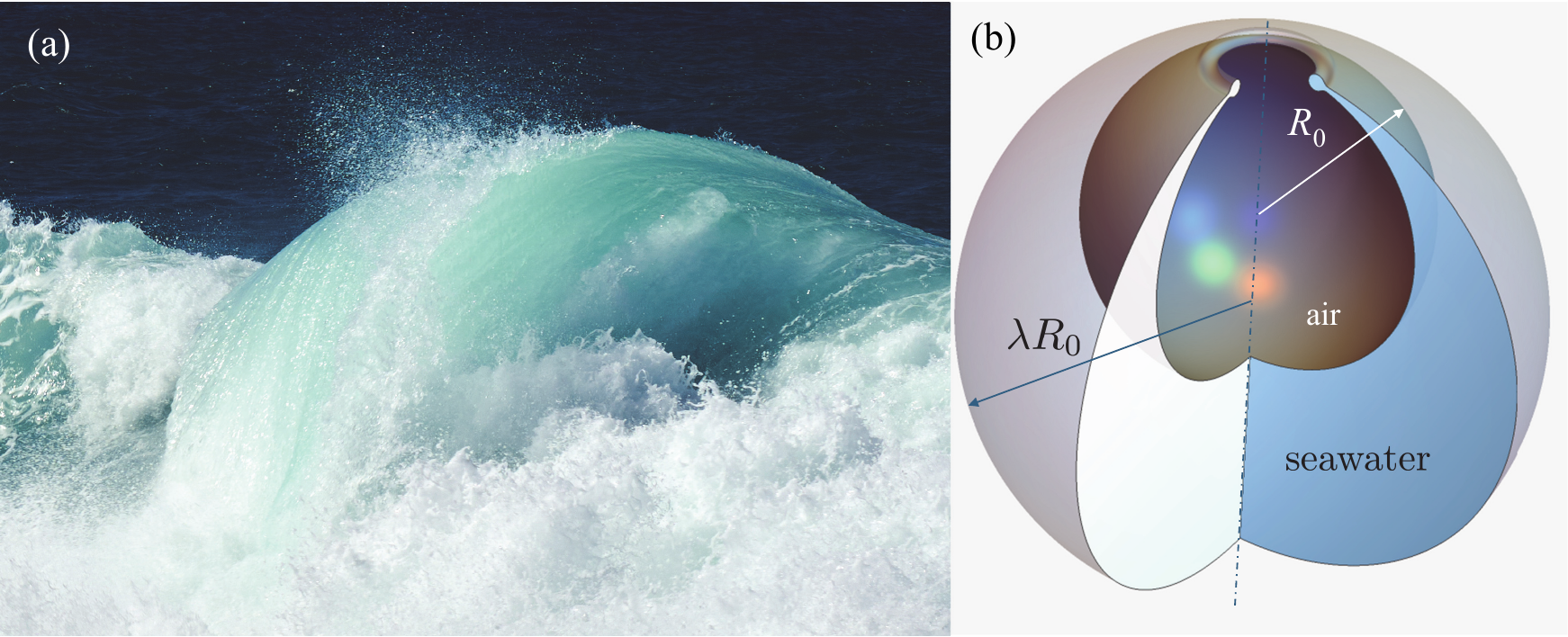}
  \caption{(\textit{a}) The crest of a spilling breaker; field of view
  $\simeq4$\,m. The whiteness of the foreground is multiple scattering at
  innumerable menisci; the photograph resolves down to about a centimetre,
  while the bubble population continues three decades below
  \citep{DeanStok2002,RiviereEtal2021}. Photograph by Juan Luis Varela.
  (\textit{b}) The configuration studied, cut away shortly after the puncture
  of the film.}
  \label{fig:context}
\end{figure*}

Bubbles bursting at the sea surface are the main source of marine aerosol
\citep{Blanchard1989,Spiel1994,DuchEtal2002,LhuiVill2012,WangEtal2017,GananCalvo2017,DeikeReichlPaulot2022}.
Almost everything we know about the process comes from bubbles bursting at
the flat surface of a deep liquid, and the sea surface is rarely like that. A
breaking wave injects a plume of bubbles spanning several decades in size
\citep{DeanStok2002,RiviereEtal2021,DerakhtiEtal2024}, and the drops torn from
the plume carry part of that gas with them \citep{Veron2015}. The crest in
figure \ref{fig:context}(\textit{a}) shows three states side by side: droplets
already in flight, a face that is still coherent but streaked with entrained
air and, in the foreground, not a surface at all but a mixture of liquid and
gas.

A drop torn from such a mixture is unlikely to be free of bubbles. Consider a
plume of void fraction $\alpha$, a detached parcel of radius $a$ and the Hinze
scale $r_H\simeq1$\,mm. The bubble size distribution peaks in volume near
$r_H$ and grows as $r^{-3/2}$ below it \citep{DeanStok2002}, so that the gas
volume sits in bubbles close to $r_H$ while their number is dominated by the
small ones. Integrating the distribution gives about $10\,\alpha\,(a/r_H)^3$
bubbles per parcel, most of the prefactor coming from the small end of the
spectrum, and this is of order one or more for millimetric parcels when
$\alpha$ is of order $0.1$. Drops that contain a bubble (hollow drops, sketched in figure
\ref{fig:context}(\textit{b})) are therefore a common product of wave breaking.
When the bubble bursts, its cap opens into the air just as it would at a flat
surface, and the droplets are ejected into open air. The difference is the
amount of liquid behind the cap, which is now finite.

The bursting of a bubble at a flat surface has been reviewed from several
angles: its role in the exchange between ocean and atmosphere
\citep{Veron2015,Deike2022ARFM}, the physics of film and jet drops and its
open questions \citep{VillermauxWangDeike2022}, bubbles in carbonated drinks
\citep{SeonLigerBelair2017}, and the whole sequence from the rise of the
bubble to the release of its droplets \citep{YuEtal2023}. For jet drops the
picture is now quantitative. Capillary waves travelling down the cavity focus
on the axis \citep{DuchEtal2002,WallsEtal2015,GananCalvo2018,GordilloRodriguez2019},
bringing the collapse close to a finite-time singularity
\citep{LaiEggersDeike2018}. The size and speed of the first droplet follow
scaling laws in the Laplace and Bond numbers
\citep{GananCalvo2017,Brasz2018,Deike2018,GananCalvoLopezHerrera2021}; no
droplets are ejected above a threshold Ohnesorge number
\citep{WallsEtal2015,GananCalvo2017}; and ensembles of simulations have given
the number and sizes of all the droplets that a single bubble produces
\citep{Berny2020,Berny2021,Berny2022}.

Aerosol size distributions, for their part, are measured at sea and in
laboratory tanks, with plunging jets, bubble plumes and wind--wave simulators
\citep{deLeeuwEtal2011,QuinnEtal2017,LeibenspergerEtal2026}. They show a
dominant accumulation mode at dry diameters between $0.1$ and $0.2\,\mu$m,
usually attributed to film drops, and how much of the submicrometre aerosol
comes from film drops and how much from jet drops is still debated
\citep{WangEtal2017,JiangEtal2022,GananCalvo2023,GHL2026,WangEtal2026}.
\citet{DeikeReichlPaulot2022} showed how to build the spectrum from the
physics of single events, by combining a per-event law
\citep{GananCalvo2017,Berny2021}, a bubble population \citep{DeanStok2002}, a criterion to separate jet from film drops \citep{LhuiVill2012}
and an ejection threshold \citep{WallsEtal2015,GhabacheSeon2016}.

Here we follow the same route for the geometry that breaking waves actually
produce: a bubble enclosed in a drop in flight. That confinement can extend
ejection below the flat-surface threshold has been shown experimentally, at
the same time as the present work, for bubbles bursting in sessile drops
\citep{GutierrezEtal2026}. Their geometry involves a pinned contact line, a
finite Bond number and four independent parameters. In a free drop the
confinement is described by a single geometric parameter and gravity plays
no role, so that the ejection boundary is a curve rather than a cloud of
points. Once measured, that curve turns out to organize all the other results
of this work.

We follow one bursting event from beginning to end, and the paper follows the
same order. After the shell is punctured, its collapse turns out to be that of
a thread of gas rather than of a cavity (\S\ref{sec:collapse}). The collapse
either launches a jet that ejects droplets or it does not, and the boundary
between the two defines a distance $\delta$ to the threshold
(\S\ref{sec:bound}). This distance controls what is ejected
(\S\ref{sec:ejected}): the volume, which follows a law with a single constant;
the first droplet, whose size goes through a minimum; and the largest
droplet, whose size is set by the cavity. Between the first and the largest
lies the full census of droplets produced by the event. Two lengths are
usually invoked to bound it. The upper one, we find, is geometric: the same
ceiling that limits the ejected volume. The lower one is the viscocapillary
length $\ell_\mu=\mu^2/\rho\sigma$, which appears explicitly in recent
parametrizations (for instance as a cut-off $x_1=0.1\,\ell_\mu/R_0$ in the
per-bubble distribution of \citet{GHL2026}) and implicitly whenever a computed
spectrum is said to end near $\ell_\mu$. Whether this lower bound belongs to
the fluid or to the finest scale resolved by the computation matters. If it is
physical, each event produces a finite number of droplets and a spray flux
can be closed by number. If it is not, every number-weighted quantity is only
a lower bound, and closure must rely on the ejected mass instead. We show that
the second alternative holds (\S\ref{sec:census}), and we carry the per-event
law over a population of bubbles to a spectrum in closed form
(\S\ref{sec:spray}). Because the smallest scales of the event approach the
cell size of the adaptive volume-of-fluid computations \citep{Popinet2009},
the refinement criterion affects some of the results; we state which ones in
\S\ref{sec:scheme}.

Finally, the event reproduces its own starting point. Some of the fragments
it produces are themselves hollow droplets, a small cavity inside a small
shell (\S\ref{sec:discussion}), which can in turn burst.

\section{Formulation and numerical method}\label{sec:form}

\subsection{Configuration}\label{sec:config}

At $t=0$ a spherical gas bubble of radius $R_0$ lies tangent internally to a
spherical liquid drop of radius $R_1=\lambda R_0$, with $\lambda>1$, and both
fluids are at rest. The liquid has density $\rho$, dynamic viscosity $\mu$ and
surface tension $\sigma$; the gas has $10^{-3}$ times the density and
$10^{-2}$ times the viscosity of the liquid. Gravity is omitted from the
computation, the Bond number being $10^{-3}$ or smaller for the drop sizes at
issue, so that both interfaces are spherical and buoyancy is negligible over
the capillary time of the event.

Lengths are scaled with $R_0$, velocities with the capillary velocity
$v_\sigma=(\sigma/\rho R_0)^{1/2}$ and times with the capillary time
$t_\sigma=(\rho R_0^3/\sigma)^{1/2}$; all variables are dimensionless
hereafter. Two groups control the problem,
\begin{equation}
  \Ohn=\frac{\mu}{(\rho\sigma R_0)^{1/2}},
  \qquad
  \Lambda=\lambda^3-1=\frac{V_{\rm liq}}{V_{\rm gas}},
  \label{eq:params}
\end{equation}
with $V_{\rm gas}=\tfrac43\pi R_0^3$ and $V_{\rm liq}=\tfrac43\pi\Lambda R_0^3$.
Since $\ell_\mu/R_0=\Ohn^2$, the viscocapillary length in units of the bubble
is known in every run. A third group appears in thin shells: the Ohnesorge
number built on the mean shell thickness $\hat h=\lambda-1$,
\begin{equation}
  \Ohn_s=\frac{\Ohn}{\hat h^{1/2}},
  \label{eq:ohs}
\end{equation}
which measures viscous damping across the shell rather than across the bubble. The
difference of radii $\hat h=\lambda-1=\Lambda/3-\Lambda^2/9+O(\Lambda^3)$ agrees
to first order with the volume-mean thickness $V_{\rm liq}/4\pi R_0^2=\Lambda/3$,
which is exact and independent of how the shell is distributed; we use
$\hat h=\lambda-1$ throughout.

\subsection{Why tangency: the modal family of initial conditions}
\label{sec:modal}

Tangency is not an assumption but the state that flight in air selects, and
the reason is best seen by writing the shell thickness as a Legendre series,
\begin{equation}
  \frac{h(\theta)}{\hat h}=1+a_1P_1(\cos\theta)+a_2P_2(\cos\theta)+\dots,
  \label{eq:modal}
\end{equation}
with $\theta$ the polar angle. The surface stiffness of mode $n$ is
$(n-1)(n+2)$, the Rayleigh spectrum of a sphere under tension. Mode $n=0$ is
not a mode but the base state, whose amplitude $\hat h$ is fixed by $\Lambda$.
Mode $n=1$ has zero stiffness: it is a rigid translation of the bubble within
the drop, and at small Bond number it costs no surface energy until the bubble
reaches the outer surface. Buoyancy, however weak, therefore moves the bubble
unopposed until it stops on contact, and the configuration with $a_1=-1$
(tangency at one pole, thickness $2\hat h$ at the other) is reached for
free. The rise time across a drop, $10^{-2}$ to $1$\,s over the sizes at
issue, is short against residence times in the air, and a bubble that never
reaches the outer surface never bursts.

Mode $n=2$ is the cheapest mode that costs anything, and it is the first with
tangency at \emph{both} poles: since $P_n(1)=1$ and $P_n(-1)=(-1)^n$, even
modes produce double tangency and odd modes single. Requiring $h\ge0$ with
$a_1$ and $a_2$ alone, and $a_2\le0$, gives the admissible triangle
$a_2\in[-1,0]$, $|a_1|\le1+a_2$, whose vertices $(\pm1,0)$ are the single
tangency studied throughout and whose apex $(0,-1)$ is double tangency, the
shell then being punctured at both poles at once. The two configurations are
the two topologies of \S\ref{sec:collapse}: one mouth, and two.

\subsection{Initial condition and numerical scheme}\label{sec:scheme}

\begin{figure}
  \centering
  \includegraphics[width=\columnwidth]{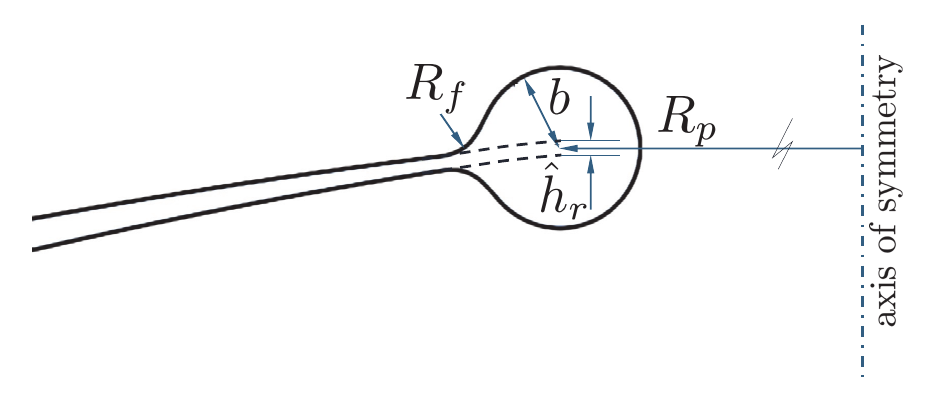}
  \caption{The initial rim in the meridional plane, defining the four lengths
  $\hat h_r$, $R_p$, $b$ and $R_f$ of \S\ref{sec:scheme}.}
  \label{fig:rim}
\end{figure}

At $t=0$ the film separating the bubble from the atmosphere is removed over a
circular patch of radius $R_p$ centred on the symmetry axis, and the opened
edge is capped by a Taylor--Culick rim. Figure \ref{fig:rim} defines the four
lengths involved. The film has dimensionless thickness $\hat h_r$ at the edge
of the hole; the rim is a torus of tube radius $b$, fixed by conservation of
the volume of film removed; and the torus is blended into the film over a
fillet of radius $R_f$.

The puncture radius is not a free parameter. The film must be resolved by the
mesh, which requires
\begin{equation}
  \hat h_r=\max\!\left(\hat h_r^{\rm phys},\,2.5\Delta\right),
  \qquad
  R_p=\left[\frac{2\hat h_r}{1-1/\lambda}\right]^{1/2},
  \label{eq:puncture}
\end{equation}
where $\Delta$ is the smallest cell size, defined below, and
$\hat h_r^{\rm phys}=1.875\times10^{-3}$ is the thickness at which the van der
Waals contribution to the disjoining pressure ruptures a clean aqueous film.
The liquid is taken to be clean throughout. The second expression follows
from the gap between the two spherical surfaces, $(r^2/2)(1-1/\lambda)$ at a
distance $r$ from the axis, and reduces to the flat-surface result
$R_p=(2\hat h_r)^{1/2}$ as $\lambda\to\infty$. The second argument of the
maximum governs on coarse meshes, so $R_p$ grows with the cell size; we have
verified insensitivity of the results to $R_p$ over the range $0.11$ to
$0.43$, and the two thinnest columns lie at or beyond its upper end
(\S\ref{sec:discussion}).

The equations are solved with Basilisk \citep{Popinet2003,Popinet2009}, using
a volume-of-fluid interface, a balanced-force surface-tension scheme and
quadtree adaptive refinement, in a square box of side $20\lambda R_0$ centred
on the drop, whose boundary lies ten drop radii from the centre of the drop
and does not influence the motion: the longest ligament recorded reaches
$10.4\,R_0$ from the centre at its first pinch-off, short of the boundary of
its box. With $N$ the maximum refinement level the smallest cell is
\begin{equation}
  \Delta=\frac{20\lambda R_0}{2^{N}} .
  \label{eq:delta}
\end{equation}
Since $\Delta$ carries $\lambda$, the same level means different physical
resolutions at different $\Lambda$, and $N$ and $\Delta$ are read from the
header of each run rather than from its label. A run terminates at a time $0.5$
after the last resolved droplet birth, or at a cap $t=4$; a case that never
ejects runs to the cap, so reaching it is the signature of no ejection.

The grid is refined on the volume fraction and on the two components of the
velocity, with wavelet tolerances $10^{-3}$, $10^{-2}$ and $10^{-2}$; the
curvature of the interface is not among the criteria. \citet{GutierrezEtal2026}
add it, with a tolerance of $10^{-8}$ that drives the mesh to the maximum level
along the whole interface. In a companion computation of one of their sessile
cases at $N=14$, adding it advances the pinch-off of the tip droplet from
$t=0.58$ to $0.46$--$0.50$ and roughly halves its radius, from $0.158$ to
$0.083$--$0.093\,R_0$, closer to their measurement, and raises the number of
escaped droplets from five to seven, while the total ejected volume is
unchanged ($0.053$) and the choice of initial rim is immaterial. Volumes
integrated over the event are therefore robust to the criterion; the size and
birth time of the first droplet, and the count, are not. The radii of the
first droplet quoted in \S\ref{sec:crest} are to be read as upper bounds at a
given level, as the drift between levels already indicates, whereas the
locations of its minimum rest on differences between neighbouring cases
computed alike.

\subsection{Droplet census, ejected mass and the first droplet}
\label{sec:census-def}

Connected liquid fragments are identified at every output time and linked
into trajectories by predicted position and conserved volume, so that each
droplet is counted once rather than once per output. Every unique droplet
carries a birth time, an equivalent radius $R_d=(3V/4\pi)^{1/3}$ with $V$ its
volume, a velocity history, and a fate: escape from the neighbourhood of the
parent drop, recoalescence with it, or survival in flight when the run ends.
The ejected mass $M_e$ is the summed volume of the droplets that escape, taken
as those with sustained outward velocity. The resolved census is the set of
unique droplets with $R_d\ge3\Delta$, the factor three being the smallest
radius at which a spherical fragment spans more than a few cells; it depends
on $N$ by construction, which is the point of \S\ref{sec:census}.

Two conventions fix what is meant by \emph{the first droplet}, and both
matter. The first droplet is the one at the tip of the train, the escaped
droplet of greatest final axial position. It is not the earliest born: a
satellite of a forked pinch-off \citep{Berny2022} can be born ahead of it.
Nor is it the largest: subsequent drops may be smaller or larger
\citep{Berny2021}. It is the fastest, the one the scaling laws describe
\citep{GananCalvo2017}, and the one a camera would point to. Its radius is the
median over the part of its trajectory in which its volume is stationary
(after the head of the ligament has consolidated and before any in-flight
coalescence), because a merger midway through the flight would otherwise
place the median between two radii, neither of them emitted.

\subsection{The campaign and its replicas}\label{sec:campaign}

The campaign comprises $285$ realizations at levels $N=12$ to $15$: $222$ at
single tangency, with
\[\Lambda\in\{1/16,1/4,1,2,4,8,16,64,512\}\] and $\Ohn$ from $0.005$ to $0.11$,
two further columns, $\Lambda=1/32$ and $1/8$, entering only the collapse
clock and energy of \S\S\ref{sec:clock}--\ref{sec:energy}; and $63$ at double
tangency (\S\ref{sec:twomouths}). Every run is kept, whatever its outcome, and
each result draws on the subset fit for it, stated where the result is quoted
and collected in appendix \ref{app:campaign}. Of the single-tangency runs,
$67$ did not terminate normally (stopped, aborted, diverged or incomplete), a
property of the cost of the calculation that we report rather than conceal.

Seventeen of the physical cases were replicated: the same case, identical
binary and arguments, run with a different number of parallel subdomains. The
code is deterministic, so a replica with the same decomposition reproduces a
run bit for bit; a different decomposition changes the order of the
floating-point reductions and hence the last bit, and the pinch-off amplifies
that difference into a divergent trajectory. The resulting spread is an
ensemble over microscopic perturbations, of the kind introduced for bubble
bursting by \citet{Berny2022} through an explicit noise in the initial
velocity; its physical analogue is real, two nominally identical bubbles
differing by vibration, thermal noise and initial shape. Replicas of a case
are collapsed to one point before any fit, their dispersion serving as its
error bar; figures show every realization. The dispersion depends strongly on
the quantity: $0.6$ to $2\,\%$ in the ejected mass, $11$ to $13\,\%$ in the
radius of the first droplet, up to $50\,\%$ in its velocity (appendix
\ref{app:replicas}).

Simulations of this kind are usually reported through maps of velocity,
pressure or vorticity at selected instants. We leave such fields aside except
where they settle a mechanism. The object of this work is the aerosol an
event ejects (its amount, its size distribution and its global consequences),
and the fields on the way there are intermediate results. We therefore state
the results in global quantities, such as the ejected volume, the droplet
census or the kinetic energy of the collapse, and in singular ones, such as
the first droplet, the instant of collapse or the location of the ejection
boundary. Each of them can be measured in every run and compared across the
whole campaign. Three-dimensional renderings of the interface, obtained by
revolving the computed profiles, are shown where the geometry of the event is
itself the point (figures \ref{fig:marginal}, \ref{fig:n2render} and
\ref{fig:n2delay}).

Tables \ref{tab:scales} and \ref{tab:columns} collect the quantities used
throughout. The first gives the definitions, with their values for water and
seawater, including the two locations of the minimum of the first droplet
established in \S\ref{sec:crest}; the second gives, column by column, the
geometry, the boundary (\ref{eq:law}), the smallest cell at the levels used,
and the bubble sizes the boundary admits in seawater.

\begin{table*}
  \begin{center}
  \small
  \begin{tabular}{llcc}
  Quantity & Definition & Water & Seawater\\[3pt]
  Ohnesorge number & $\Ohn=\mu/(\rho\sigma R_0)^{1/2}$ & -- & --\\
  Viscocapillary length & $\ell_\mu=\mu^2/\rho\sigma=\Ohn^2R_0$ & $13.8$\,nm & $15.5$\,nm\\
  Thermal length & $\ell_T=(k_BT/\sigma)^{1/2}$ & $0.24$\,nm & $0.24$\,nm\\
  Flat-surface threshold & $\Ohn_c=0.043$;\ \ $R_{0,c}=\ell_\mu/\Ohn_c^2$ & $7.5\,\mu$m & $8.4\,\mu$m\\
  Boundary & $\Ohn_1=\Ohn_c(1+2\beta/\lambda)$, $\beta=0.831$ & -- & --\\
  \quad smallest ejecting bubble, $\Lambda=1$ & $R_{0,\min}=\ell_\mu/\Ohn_1^2$ & $1.4\,\mu$m & $1.6\,\mu$m\\
  Distance to the boundary & $\delta=1-\Ohn/\Ohn_1$ & -- & --\\
  \quad minimum of $R_{d1}/\ell_\mu$ & $\delta^*=0.21\pm0.03$ & -- & --\\
  \quad minimum of $R_{d1}/R_0$ & $\delta^*_R\simeq0.30\pm0.03$ & -- & --\\
  \quad bubble at $\delta^*$, $\Lambda=16$ & $R_0=\ell_\mu/[\Ohn_1(1-\delta^*)]^2$ & $4.4\,\mu$m & $5.0\,\mu$m\\
  Shell thickness & $\hat h=\lambda-1$, $\lambda=(1+\Lambda)^{1/3}$ & -- & --\\
  Smallest cell & $\Delta=20\lambda R_0/2^N$ & \multicolumn{2}{c}{table \ref{tab:columns}}\\
  \end{tabular}
  \caption{Definitions and scales. Fluid properties at $20\,^\circ$C, with
  salinity $35$\,g\,kg$^{-1}$ for seawater \citep{Sharqawy2010}: $\mu=1.00$ and
  $1.08$\,mPa\,s, $\rho=998$ and $1025$\,kg\,m$^{-3}$, $\sigma=72.8$ and
  $73.5$\,mN\,m$^{-1}$. $\ell_\mu$ grows by about two thirds from $20$ to
  $10\,^\circ$C, and every length derived from it with it.}
  \label{tab:scales}
  \end{center}
\end{table*}

\begin{table*}
  \begin{center}
  \small
  \def~{\hphantom{0}}
  \begin{tabular}{lcccccccc}
  & & & & & \multicolumn{2}{c}{$10^3\Delta/R_0$} & & $R_{0,\min}$\\
  $\Lambda$ & $\lambda$ & $\hat h$ & $2/\lambda$ & $\Ohn_1$ &
  $N=13$ & $N=15$ & $N$ used & ($\mu$m)\\[3pt]
  $1/16$ & $1.020$ & $0.020$ & $1.960$ & $0.113^\dagger$ & $2.49$ & $0.62$ & $14$--$15$ & $1.2^\dagger$\\
  $1/4$  & $1.077$ & $0.077$ & $1.857$ & $0.109$ & $2.63$ & $0.66$ & $13$--$15$ & $1.3$\\
  $1$    & $1.260$ & $0.260$ & $1.587$ & $0.100$ & $3.08$ & $0.77$ & $12$--$15$ & $1.6$\\
  $2$    & $1.442$ & $0.442$ & $1.387$ & $0.093$ & $3.52$ & $0.88$ & $13$, $15$ & $1.8$\\
  $4$    & $1.710$ & $0.710$ & $1.170$ & $0.085$ & $4.17$ & $1.04$ & $12$--$15$ & $2.2$\\
  $8$    & $2.080$ & $1.080$ & $0.962$ & $0.077$ & $5.08$ & $1.27$ & $13$, $15$ & $2.6$\\
  $16$   & $2.571$ & $1.571$ & $0.778$ & $0.071$ & $6.28$ & $1.57$ & $12$--$15$ & $3.1$\\
  $64$   & $4.021$ & $3.021$ & $0.497$ & $0.061$ & $9.82$ & $2.45$ & $12$--$13$ & $4.2$\\
  $512$  & $8.005$ & $7.005$ & $0.250$ & $0.052$ & $19.5$~ & $4.89$ & $15$ & $5.8$\\
  $\infty$ & -- & -- & $0$ & $0.043$ & -- & -- & -- & $8.4$\\
  \end{tabular}
  \caption{The columns of the campaign. $\Ohn_1$ is the boundary (\ref{eq:law})
  with $\Ohn_c=0.043$ and $\beta=0.831$; $R_{0,\min}=\ell_\mu/\Ohn_1^2$ is the
  smallest bubble that ejects in seawater. The viscocapillary length is
  resolved, $\ell_\mu\ge\Delta$, for $\Ohn\ge(\Delta/R_0)^{1/2}$: $0.025$ to
  $0.070$ at $N=15$ from the thinnest to the thickest column.
  $^\dagger$Law value in the thin-shell column excluded from the fit, where
  the boundary is not closed (\S\ref{sec:bound}).}
  \label{tab:columns}
  \end{center}
\end{table*}

\section{The collapse: one mechanism, two topologies}\label{sec:collapse}

\subsection{A thread of gas, pinched next to its mouth}\label{sec:thread}

\begin{figure*}
  \centering
  \includegraphics[width=\textwidth]{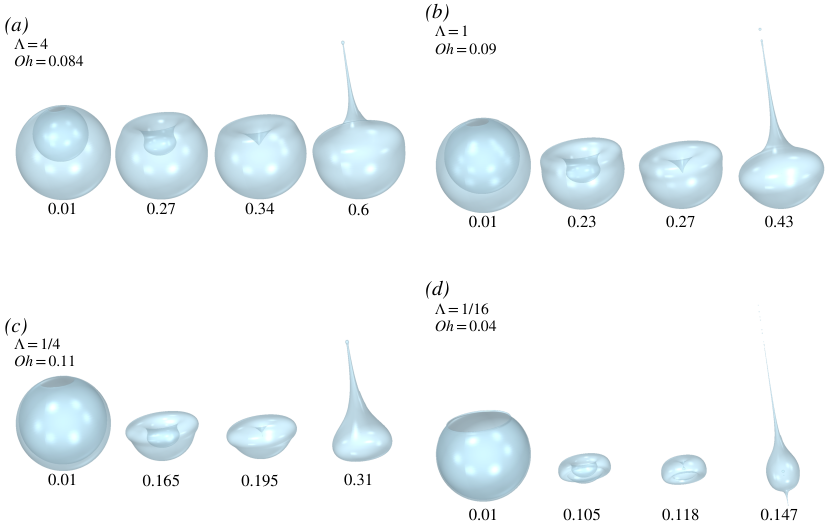}
  \caption{Marginal emission at $(\Lambda,\Ohn)=$ (\textit{a}) $(4,0.084)$,
  (\textit{b}) $(1,0.09)$, (\textit{c}) $(1/4,0.11)$ and (\textit{d})
  $(1/16,0.04)$, four instants each, with the dimensionless time below every
  frame. Ejection succeeds in (\textit{a}); in (\textit{b}) collisions within
  the train decide the fate of individual droplets (at $N=15$, shown, four to
  seven escape depending on the realization, and none at $N=12$); in
  (\textit{c}) ejection fails by refilling of the neck.
  Panel (\textit{d}) is the thin-shell, twin-jet regime; the retrograde jet is
  visible at the last instant.}
  \label{fig:marginal}
\end{figure*}

Once the rim has retracted, the capillary wave that runs down the cavity
wall focuses on the axis and squeezes the remaining gas into a slender
column. What collapses is therefore not a cavity but a cylindrical thread of
gas, and it pinches next to its mouth, because the thread has a point that it
cannot move: the floor of the cavity when the shell is punctured once, the
stagnation point imposed by symmetry when it is punctured at both poles. The
number of pinch-offs is the number of mouths, and the head of the ejected
ligament forms beyond each of them by reversal of curvature.
Figure \ref{fig:marginal} shows the event with one mouth in four shells,
three of them near their ejection boundary and the fourth in the
thin-shell regime: the cavity collapses into a cone whose apex reverses into
the jet.

The approach to pinch-off follows the inertial law of a gas thread. Fitting
the neck radius at fixed axial stations to $R\propto\tau^{\alpha}$, with
$\tau=t_s-t$ and the singular time $t_s$ left free at each station, gives
\begin{equation}
  \begin{gathered}
  \alpha=0.513\pm0.025\quad(\text{one mouth}),\\
  \alpha=0.486\pm0.005\quad(\text{two mouths}),
  \end{gathered}
  \label{eq:alpha}
\end{equation}
the first over four stations of the best-resolved single-tangency case
($\Ohn=0.09$, $\Lambda=1/16$, $N=15$, $\ell_\mu/\Delta=13$), the second at the
stations where the double-tangency thread actually pinches ($\Ohn=0.03$,
$\Lambda=1/2$), converged between $N=13$ and $14$ and symmetric about the
equator to five digits. Following the global minimum of the neck instead
measures the stagnation point, not the pinch, and returns a spurious
$0.66$--$0.78$. Both exponents in (\ref{eq:alpha}) are compatible with the
value $1/2$ of inviscid gas-thread collapse, whose effective slope over the
range accessible to computation is close to $0.53$
\citep[\S9.3]{EggersFontelos2015}. With two mouths the axial scale of the neck is frozen: its axial length
$\ell$ varies as $\ell\propto a^{0.21}$ with the neck radius $a$, with $N=13$ and $14$ agreeing to one per cent at equal
neck radius. The singular time grows outward from the impenetrable point, so
that the pinch travels towards the mouth; and the trapped column of gas does
not vanish but persists at $0.7$ to $2.2$ cells to the end of the run, an
incompressible gas with no way out.

\subsection{The clock of the collapse}\label{sec:clock}

\begin{figure*}
  \centering
  \includegraphics[width=\textwidth]{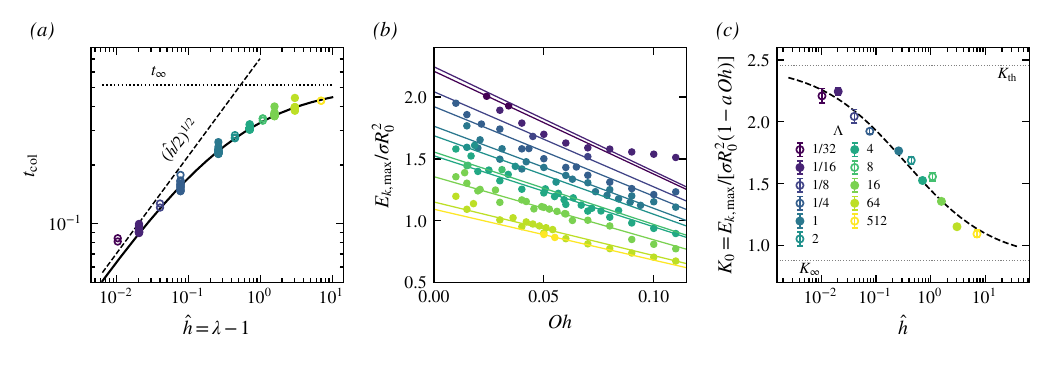}
  \caption{(\textit{a}) Collapse time against shell thickness; filled symbols,
  columns used in the fit of (\ref{eq:clock}), solid line; open symbols,
  columns not used. Dotted, $t_\infty$; dashed, Taylor--Culick
  $(\hat h/2)^{1/2}$. (\textit{b}) Maximum kinetic energy against $\Ohn$, with
  $K_0(1-a\,\Ohn)$, $a=3.8$. (\textit{c}) $K_0$ against $\hat h$; dashed, (\ref{eq:K0}); dotted, its
  limits $K_{\rm th}$ and $K_\infty$. Colours, $\Lambda$ from $1/32$ to $512$.}
  \label{fig:clock}
\end{figure*}

The instant of collapse can be read from every run, since it marks three
events in close succession: the maximum of the kinetic energy, when the wave
focuses; a cliff in the time step, when the neck pinches; and the time of
jet formation recorded by the solver, $0.08$ later. Over the columns in which
it is converged in mesh, to between $0.8$ and $2.8\,\%$ across levels, the
collapse time rises with the shell thickness and saturates:
\begin{equation}
  \begin{gathered}
  t_{\rm col}=t_\infty\,\frac{\hat h^{\,q}}{b+\hat h^{\,q}},\\
  t_\infty=0.516,\quad q=0.550,\quad b=0.564,
  \end{gathered}
  \label{eq:clock}
\end{equation}
with $R^2=0.9995$, preferred over a single power law by $7.3$ in the Akaike
information criterion; the local slopes $\mathrm{d}\ln t/\mathrm{d}\ln\hat h$ fall
monotonically from $0.37$ to $0.13$, so that no power law describes the
data. Both limits are physical. For a thick shell $t_{\rm col}\to0.52$
capillary times of the cavity, the value of the unbounded bath. For a thin
shell $t_{\rm col}\propto\hat h^{1/2}$, which is Taylor--Culick: the rim
retracts at $(2\sigma/\rho\hat h R_0)^{1/2}$ over a fixed distance, whence
$t=(\hat h/2)^{1/2}$, predicting $0.101$ at $\Lambda=1/16$ against $0.090$
measured. The crossover lies at $\hat h^*\simeq0.35$. The exponent is secure;
the thin-shell prefactor is not, because the two thinnest columns carry
puncture radii beyond the range over which insensitivity was verified
(\S\ref{sec:discussion}).
The law was fitted on five columns. Of the six that did not enter the fit,
$\Lambda=1/8$, $1/4$, $2$, $8$ and $512$ fall on it to within $4\,\%$
(figure~\ref{fig:clock}(\textit{a})), and the thinnest shell, $\Lambda=1/32$,
exceeds it by $28\,\%$, as the unsettled thin-shell prefactor allows.

\subsection{The energy is set by the cavity}\label{sec:energy}

The maximum of the kinetic energy of the liquid, in units of
$\sigma R_0^2$, depends on both parameters, and separably
(figure~\ref{fig:clock}(\textit{b,c})):
\[
  E_{k,\max}\simeq K_0(\hat h)\,(1-a\,\Ohn),\qquad a=3.8\pm0.1,
\]
with an rms deviation of $3.6\,\%$ over 119 cases from $\Lambda=1/32$ to
$512$. The coefficient $a$ is a first-order viscous correction of order
unity, specific to the spherical hollow drop; in other geometries, such as a
bubble in a sessile drop, it would be different and would depend on the Bond
number. The general result is $K_0$: it falls monotonically from $2.25$ for
the thinnest shell to $1.09$ for $\Lambda=512$, along a sigmoid in
$\ln\hat h$ that is the collapse clock itself,
\begin{equation}
  K_0\simeq K_{\rm th}-(K_{\rm th}-K_\infty)\,\frac{t_{\rm col}}{t_\infty}
  =K_{\rm th}-(K_{\rm th}-K_\infty)\,\frac{\hat h^{\,q}}{b+\hat h^{\,q}},
  \label{eq:K0}
\end{equation}
with $K_{\rm th}=2.46\pm0.05$, $K_\infty=0.88\pm0.05$, $q$ and $b$ those of
(\ref{eq:clock}), and an rms deviation of $0.063$
over the eleven columns ($\hat h$ from $0.010$ to $7.0$). The two limits are
finite, the thin sheet and the cavity under a bath, and the energy is of order
unity over a factor of sixteen thousand in the volume of liquid. The energy that the collapse deposits is therefore set by
the cavity to within a factor of two, not by the liquid, and it is deposited
on a mass that grows as $\Lambda$; a thin shell receives more energy per unit
mass on both counts. That is the measured content of the statement, used in
\S\ref{sec:mass}, that the gas volume is the template of the ejected mass
rather than a reservoir.

\subsection{The emission law}\label{sec:emission}

The collapse ends in a reversal of curvature on the axis: a parabolic profile
that degenerates into a nipple, the proto-jet \citep[figure~2]{GananCalvo2017},
the two stages following each other within $10^{-4}$ capillary times in both
topologies. Defining the
tip as the furthest point of the interface within $1.5\Delta$ of the axis and
its radius of curvature $R$ from the twelve cells behind it, the tip velocity
$v$ obeys, along the whole trajectory from its formation to the first
pinch-off,
\begin{equation}
  v^2R=k,\qquad\text{i.e.}\qquad
  \frac{R}{\ell_\mu}=k\left(\frac{v}{v_\mu}\right)^{-2},
  \label{eq:emission}
\end{equation}
with $v_\mu=\sigma/\mu$: $v R^{1/2}=5.26\pm0.15$ at five instants for one mouth
($k=27.7$) and $3.06\pm0.18$ for two ($k=9.3$), each over a factor $2.7$ in $R$.
This is regime (ii) of \citet{GananCalvoLopezHerrera2021}; their regime (i),
$vR=\text{const}$, was adopted by \citet{GordilloBlancoRodriguez2023}
from a potential-flow theory of cavity collapse. Regime (ii) is measured here not at one
instant but along the whole flight of the tip. The component exponents are
mutually consistent, $x\propto\tau^{0.68}$, $R\propto\tau^{0.56}$,
$v\propto\tau^{-0.28}$, the first close to the $2/3$ of an inertial--capillary
wedge. Over the same instants the tip travels at $16$ to $10$ times the
viscocapillary velocity with one mouth and $5$ to $3$ times with two, an inertial--capillary regime; and yet
its radius stays below $\ell_\mu$ ($0.11$ to $0.28\,\ell_\mu$ with one mouth,
$0.32$ to $0.90$ with two), so that the two usual criteria of regime,
velocity and size, decouple in the confined event.

Two cautions attach to these numbers. With one mouth the tip radius is $1.4$
to $3.7$ cells, and its monotonic variation over a factor $2.7$ argues against
anchoring to the mesh without excluding it; with two it is $0.5$ to $1.3$
cells, so that $k=9.3$ is below the grid and is quoted as indicative only. The
coincidence $k/\Ohn\simeq310$ in both cases rests on two points and three
candidate parameters, and we do not use it.

Near threshold the neck freezes while the ligament keeps lengthening (at
$0.41\,\ell_\mu$ with one mouth and $1.81\,\ell_\mu$ with two, with the
ligament growing by factors $2.5$ and $3.2$). A stretched ligament can be
frozen by its own stretching: the relevant measure is the ratio
$\dot\varepsilon\tau$ of the stretching rate to the capillary rate
\citep{MarmottantVillermaux2004}, and its local value of order unity marks the
end of stretching control at the tip of a Worthington jet in
\citet{GordilloGekle2010}, whose stretching Weber number is its square. That is
not what happens here. Where the neck is stretched, with
$\dot\varepsilon=\partial u_x/\partial x$ measured on the axis, the ratio stays
small: $\dot\varepsilon\,t_p\le0.23$ ($0.03$ to $0.12$ with one mouth), with
$t_p=\sqrt8\,(\rho a^3/\sigma)^{1/2}+6\mu a/\sigma$ the inverse growth rate of
the fastest capillary mode of a thread of the neck radius $a$ in Weber's
long-wave theory \citep{Weber1931}, which contains Rayleigh's inviscid and
viscous limits; both terms matter, since the local Ohnesorge number
$\Ohn(R_0/a)^{1/2}$ lies between $0.7$ and $1.6$. Once frozen, the neck is not stretched at all but fed:
$\dot\varepsilon$ turns negative, through a stagnation zone like the one that
delays the inertial--viscous regime in the thinning of filaments
\citep{CastrejonPitaEtal2015}. The frozen neck is a cylinder in Laplace
equilibrium with the local pressure, $pa/\sigma=1.00\pm0.02$ with one mouth and
$0.97\pm0.03$ with two over sixteen instants: it thickens as the ligament
depressurizes, which accounts for the two arrest radii without reference to
$\ell_\mu$. Stretching decides little here, and the ejection is decided by
energy (\S\ref{sec:twomouths}).

\subsection{The second topology: double tangency}\label{sec:twomouths}

\begin{figure*}
  \centering
  \includegraphics[width=\textwidth]{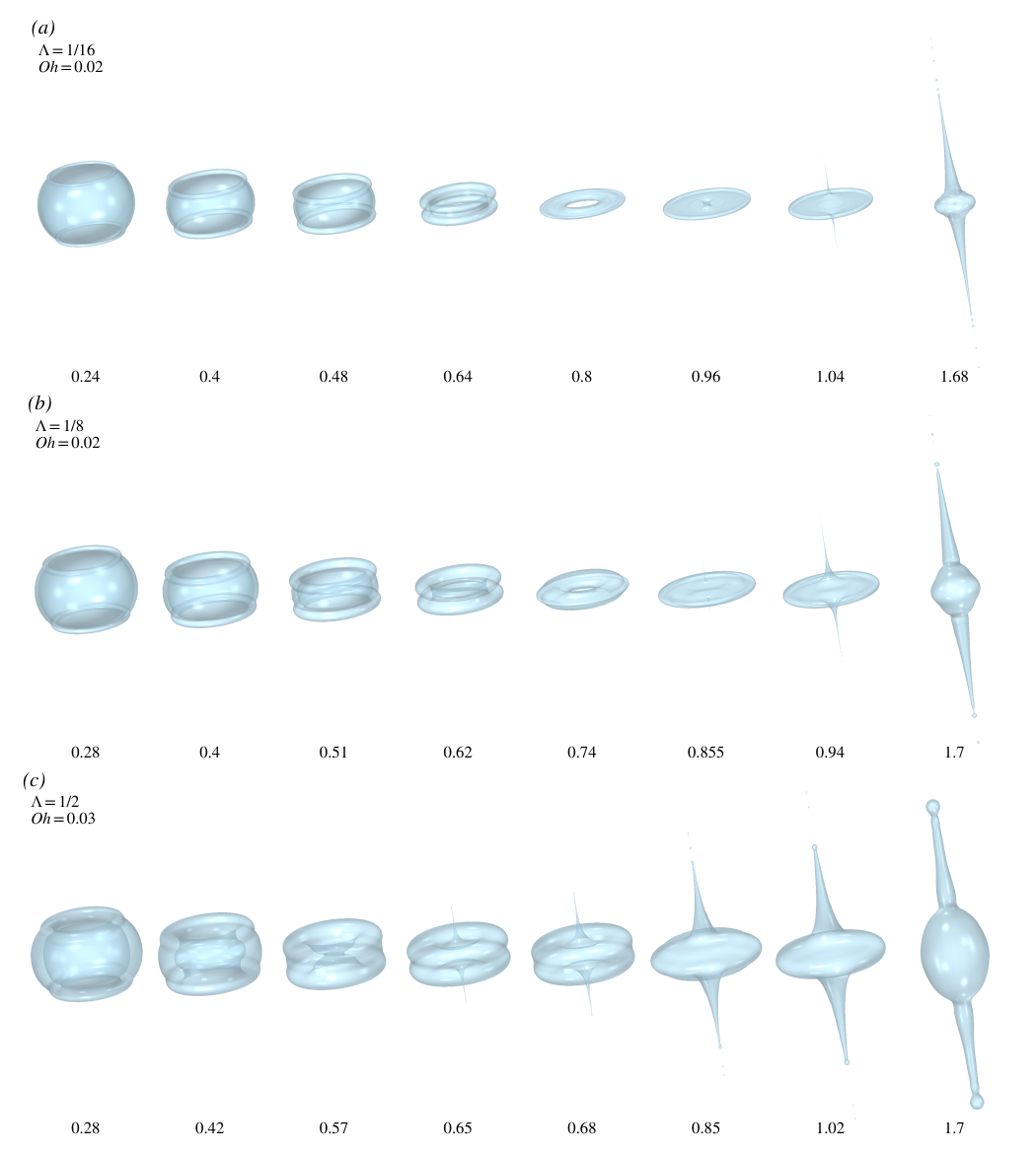}
  \caption{Double tangency with simultaneous punctures ($\Delta t=0$): the
  computed interface revolved about the axis, at eight instants in each row,
  with $t/t_{\rm per}$ below every frame. (\textit{a}) $\Lambda=1/16$,
  $\Ohn=0.02$; (\textit{b}) $\Lambda=1/8$, $\Ohn=0.02$; (\textit{c})
  $\Lambda=1/2$, $\Ohn=0.03$. The scale is common within each row.}
  \label{fig:n2render}
\end{figure*}

Double tangency (\S\ref{sec:modal}) puts the same mechanism behind two mouths,
and it can be driven asymmetrically, by opening the second pole a time
$\Delta t$ after the first. Its computations use a box of $10R_0$, since there
are no long axial jets to accommodate far from the drop, two punctures of radius
$R_p=0.35$, the thickness profile $h(\theta)=\tfrac32\hat h\sin^2\theta$ of the
$n=2$ mode, the delay imposed either as a true delay in time or, equivalently,
by opening the leading pole with a wider mouth so that its rim reaches the
equator earlier ($\cos\theta_N=\cos\theta_S-\Delta t/t_{\rm per}$, with
$t_{\rm per}$ the time the rims take to reach the equator, given below), and no mirror
symmetry imposed, so that the departure of the symmetric runs from the equator
is itself a measure of error. The film is resolved by some fifty cells at the
equator, at $N=13$ for $\Lambda=1/8$ and $N=14$ for $\Lambda=1/16$. The family
comprises $63$ realizations, $48$ of them completed, over $\Lambda=1/16$ to
$1/2$ and delays up to $0.2$ collision times. Figure \ref{fig:n2render} shows
the symmetric event in three shells. The two rims run from the poles towards
the equator, with capillary waves on the inner and outer surfaces of the
shell; in the thin shells they merge into a disc whose centre then collapses
on the axis and launches two opposed jets, while in the thicker one the axial
collapse sets in before the rims have merged.

\textit{Where the rims meet.} With $h\propto\sin^2\theta$ the Taylor--Culick
rims run along the shell at $V(\theta)=V_2/\sin\theta$,
$V_2=2/(3\hat h)^{1/2}$, and reach the equator at the peripheral time $t_{\rm per}=R_m/V_2$, with
$R_m$ the mean radius of the shell. A delay $\Delta t$ displaces the collision
to the latitude
\begin{equation}
  \cos\theta_m=-\frac{\Delta t}{2t_{\rm per}} ,
  \label{eq:collision}
\end{equation}
independent of $\Ohn$, of $\Lambda$ and of the puncture radius. The symmetric
runs meet at $\theta_m=90.000^\circ$; with $\Delta t=0.1\,t_{\rm per}$,
(\ref{eq:collision}) predicts $92.87^\circ$ and the runs give $92.04^\circ$ and
$92.64^\circ$ at $\Ohn=0.02$ and $0.05$. The difference between the two lies
within the $1^\circ$ binning of the measurement, and is not interpreted. The
hypothesis behind the law, a rim running at $V_2$, is checked directly: the
axial velocity of the rim is $5.594\pm0.051$ against $V_2=5.770$, $3\,\%$ low,
with a dispersion of $0.9\,\%$ over two values of $\Ohn$ and two delays.

\textit{The threshold carries a constant toll.} Judged event by event from the
flow fields, as for single tangency, the symmetric threshold is given in
table \ref{tab:ohc2}, with $\Ohn_1$ from (\ref{eq:law}).

\begin{table*}
  \begin{center}
  \def~{\hphantom{0}}
  \begin{tabular}{lcc}
  $\Lambda$ & $\Ohn_c^{(2)}$ & $\Ohn_c^{(2)}/\Ohn_1$\\[3pt]
  $1/16$ & $(0.030,\,0.050]$ & --\\
  $1/8$  & $0.0325\pm0.0025$ & $0.291$\\
  $1/4$  & $0.0325\pm0.0025$ & $0.297$\\
  $1/2$  & $\simeq0.0285$    & $0.270$\\
  \end{tabular}
  \caption{Threshold of the symmetric double-tangency event, judged event by
  event from the flow fields, and its ratio to the single-tangency boundary
  $\Ohn_1$ of (\ref{eq:law}).}
  \label{tab:ohc2}
  \end{center}
\end{table*}

The indirect collapse (first a meridional
collapse at the periphery, only then the thread on the axis) costs a
constant factor, $\Ohn_c^{(2)}\simeq0.29\,\Ohn_1(\lambda)$, while $\hat h$
varies by a factor seven. Two cautions keep that from being more than
compatibility: over this range $\Ohn_1$ itself varies by only $7\,\%$, and at
$\Lambda=1/2$ the neck is resolved by fewer than half a cell per $\ell_\mu$, in
the direction that would lower the threshold. At threshold a single droplet is
emitted, at $\Lambda=1/16$, $1/8$ and $1/4$ alike: the count goes to one as
$\delta\to0$, as it does for single tangency (\S\ref{sec:bound}).

\textit{What decides ejection.} One scalar orders every case: the largest
axial kinetic energy of the liquid after the rims collide. Ejection occurs
when it exceeds
\begin{equation}
  E_{x,\rm jet}\simeq(0.61\pm0.03)\,\sigma R_0^2 ,
  \label{eq:kex}
\end{equation}
over eight cases in four columns, with a single exception at $\Lambda=1/2$,
where the jet forms but its tip droplet grows almost indefinitely, always on
the point of detaching, without pinching off: an arrested pinch-off, the
same event observed at single tangency on the boundary of $\Lambda=1/4$
(figure \ref{fig:marginal}(\textit{c})). Failure takes two other forms besides:
the break-up of a thick ligament without emission, and, at $\Lambda=1/16$ and
$\Ohn=0.05$, a secondary capillary wave that swallows the incipient jet.

\begin{figure*}
  \centering
  \includegraphics[width=\textwidth]{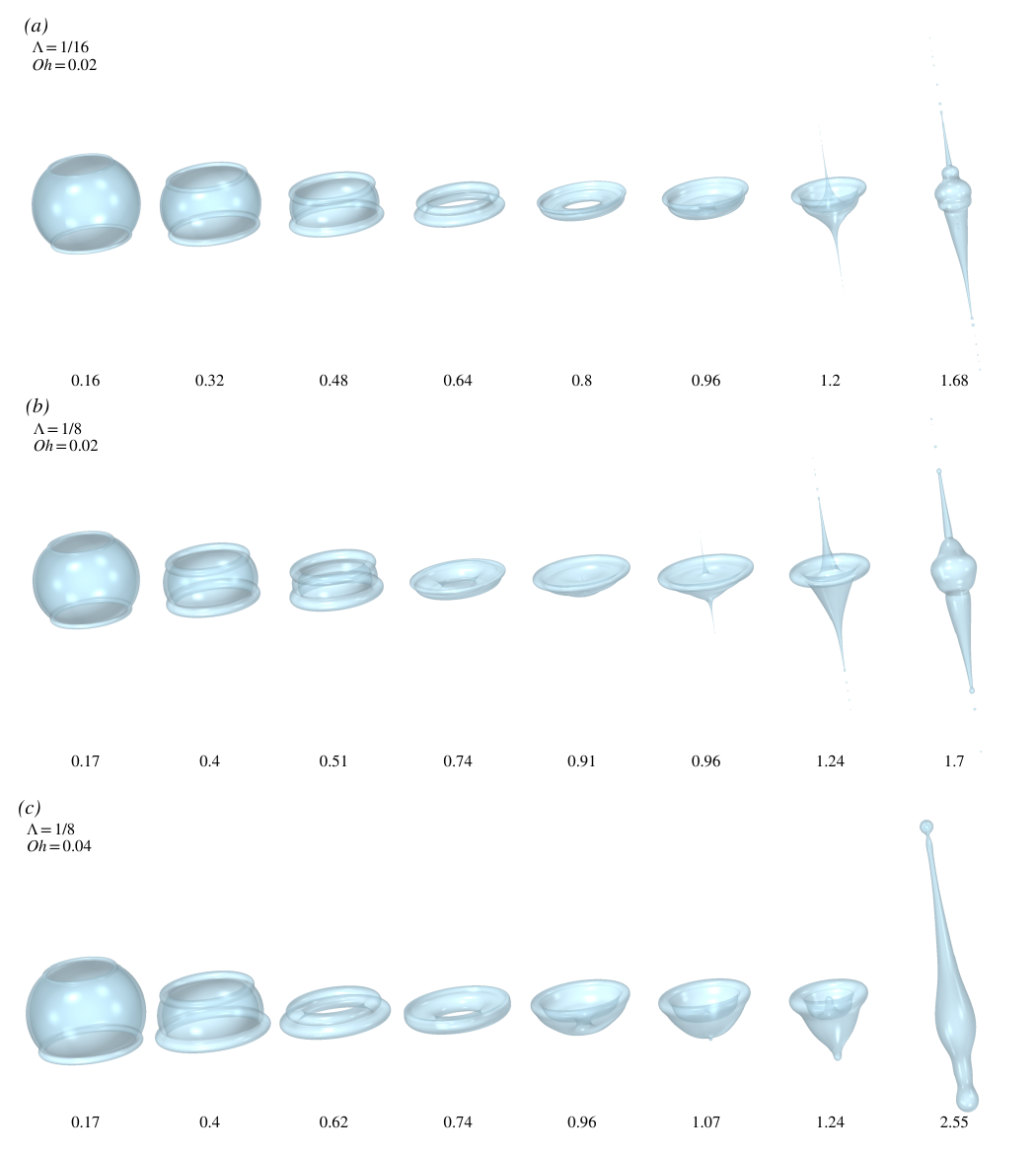}
  \caption{Double tangency with delayed puncture, rendered as in figure
  \ref{fig:n2render}, with $t/t_{\rm per}$ below every frame; the upper pole
  opens first. (\textit{a}) $\Lambda=1/16$, $\Ohn=0.02$,
  $\Delta t=0.1\,t_{\rm per}$; (\textit{b}) $\Lambda=1/8$, $\Ohn=0.02$,
  $\Delta t=0.1\,t_{\rm per}$; (\textit{c}) $\Lambda=1/8$, $\Ohn=0.04$,
  $\Delta t=0.2\,t_{\rm per}$, which does not eject.}
  \label{fig:n2delay}
\end{figure*}

\textit{Asymmetry dissipates.} The delay adds directed axial momentum
(the ratio of axial to radial kinetic energy is
$E_x/E_y=0.232+26\,(\Delta t/t_{\rm per})^2$ at the collision) instead of
sharing it between two opposed jets, and yet it suppresses ejection: at
$\Lambda=1/8$ and $\Ohn=0.030$ the symmetric event ejects and the event with
$\Delta t=0.2\,t_{\rm per}$ does not. Dissipation in the asymmetric collision
outweighs the gain in directed momentum, and the threshold falls with the
delay.
Figure \ref{fig:n2delay} shows the delayed event. With $\Delta t=0.1\,t_{\rm per}$
the rims meet south of the equator, the disc curves into a cup and the collapse
launches two jets of unequal length; with $\Delta t=0.2\,t_{\rm per}$ at
$\Ohn=0.04$ the cup folds on itself and the shell ends as a single elongated
body without ejecting.

What the second topology establishes is that the exponent of the collapse,
(\ref{eq:alpha}), and the form of the emission law, (\ref{eq:emission}), do not
depend on how many mouths the thread has: the singularity is local, and only
the efficiency with which the flow upstream feeds the neck changes between the
two. That reading makes a testable prediction, which we record without having
tested it: the minimum of the first droplet in $\Ohn$ found for one mouth
(\S\ref{sec:crest}), associated at a bath with the recirculation that forms
around the trapped bubble \citep{GananCalvoLopezHerrera2021}, should be absent
for two, where the stagnation point inhibits that recirculation.

\section{The ejection boundary}\label{sec:bound}

\subsection{Two first-order mechanisms}

Two effects shift the boundary, both of order $1/\lambda$ and both favouring
ejection. The first is an added Laplace overpressure: before puncture the gas
stands at $2(1+1/\lambda)$ above ambient, in units of $\sigma/R_0$, so that the
outer surface contributes $1/\lambda$ times as much as the bubble itself and
confinement supplies extra pressure to drive the collapse. The same mechanism
has been identified in the sessile geometry by \citet{GutierrezEtal2026}. The
second is a reduction of the inertia that the collapse must overcome: for
radial flow in a shell of outer radius $\lambda$ with a free outer surface,
the kinetic energy set in motion by a cavity of radius $1$ collapsing at a
given rate is smaller than in an unbounded bath by a factor $(1-1/\lambda)$,
because the liquid beyond $\lambda$ is not there to be moved. Both effects
being of the same order, the boundary takes the form
\begin{equation}
  \Ohn_1(\lambda)=\Ohn_c\left(1+\frac{2\beta}{\lambda}\right),
  \label{eq:law}
\end{equation}
where $\Ohn_1$ is the largest Ohnesorge number at which the event still
ejects, $\Ohn_c$ its flat-surface value and $\beta$ a coefficient of order
unity, so that $2\beta$ is the slope of $\Ohn_1/\Ohn_c$ against $2/\lambda$,
the abscissa of figure \ref{fig:map}, which is $0$ for a flat surface and $2$
for a vanishing shell.

\subsection{Measurement}

The boundary was located by bisection in $\Ohn$ at fixed $\Lambda$. The
measured values, column by column, are collected in table \ref{tab:bound}.

\begin{table*}
  \begin{center}
  \def~{\hphantom{0}}
  \begin{tabular}{lcccc}
  $\Lambda$ & $2/\lambda$ & $\Ohn_1$ (adjudicated) & $N=15$ bracket by escape &
  law (\ref{eq:law})\\[3pt]
  $1/4$ & $1.857$ & $0.108\pm0.003$   & $\ge0.09^{\,a}$   & $0.109$\\
  $1$   & $1.587$ & $0.105\pm0.005$   & $\ge0.10$         & $0.100$\\
  $4$   & $1.170$ & $0.0845\pm0.003$  & $\ge0.084$        & $0.085$\\
  $16$  & $0.778$ & $0.070\pm0.005$   & $(0.070,\,0.080)$ & $0.071$\\
  $64$  & $0.497$ & $0.060\pm0.005$   & $\ge0.054^{\,a}$  & $0.061$\\
  $512$ & $0.250$ & $0.0525\pm0.0025$ & $(0.050,\,0.055)$ & $0.052$\\
  \end{tabular}
  \caption{The ejection boundary, column by column: the value adjudicated by
  bisection, the bracket closed by the escape criterion at $N=15$, and the law
  (\ref{eq:law}) with $\Ohn_c=0.043$ and $\beta=0.831$. $^a$Lower bound from
  $N=13$.}
  \label{tab:bound}
  \end{center}
\end{table*}

\begin{figure}
  \centering
  \includegraphics[width=\columnwidth]{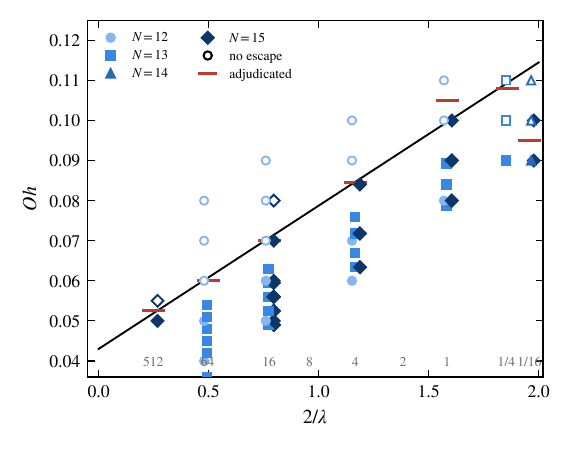}
  \caption{The ejection boundary in the plane $(2/\lambda,\Ohn)$. Every run
  near the boundary is shown at its column (labels at the bottom), offset
  slightly by level $N$: filled, at least one droplet escapes; open, none
  has escaped by $t=1.5$. Only open symbols at $N=15$ are upper bounds. Bars are
  the adjudicated $\Ohn_1$; the line is (\ref{eq:law}) with $\Ohn_c=0.043$ and
  (\ref{eq:beta}). Columns $\Lambda=2$ and $8$ carry no runs near the boundary.}
  \label{fig:map}
\end{figure}

The criterion is strict: an event ejects if at least one droplet escapes, and
an event is counted as not ejecting only if the run reaches $t\ge1.5$, because
near the boundary the first escaped droplet is born late, at up to $t=1.32$.
Since a coarse mesh assigns the less ejective regime, an escape at any level
is a lower bound on $\Ohn_1$, but only a failure at $N=15$ is an upper bound
(figure \ref{fig:map}). By that standard two columns are closed at $N=15$
(table \ref{tab:bound}),
$\Lambda=512$ in $(0.050,0.055)$ and $\Lambda=16$ in $(0.070,0.080)$, both on
the law; the others carry the adjudicated values above, whose upper bounds
come from coarser levels and are therefore provisional, and every escape
recorded at $N=15$ is consistent with them.
A free fit of (\ref{eq:law}) returns an intercept $\Ohn_c=0.0434\pm0.0026$,
which agrees to within $0.15$ standard deviations with the flat-surface
threshold $\Ohn_c\simeq0.043$ established independently at a bath
\citep{GananCalvo2017,WallsEtal2015}. Since the limit $\lambda\to\infty$ must
reproduce the bath, that agreement licenses treating $\Ohn_c$ as known, so
that only the slope is fitted:
\begin{equation}
  \beta=0.831\pm0.028 ,\qquad \chi^2/\nu=0.29 ,
  \label{eq:beta}
\end{equation}
over the six adjudicated columns, every one within its error bar of the law
for $\Lambda\ge1/4$ (figure \ref{fig:map}). The fitted law (\ref{eq:law}) is thus anchored at the
flat-surface limit and carries a single adjustable parameter.

\subsection{Saturation in thin shells}

At the thin-shell end the law saturates. As $\lambda\to1$ the relevant measure
of viscous damping is the shell Ohnesorge number (\ref{eq:ohs}). When
$\Ohn_s$ exceeds about $0.6$ the collapse is overdamped across the thickness
of the shell before a ligament can be assembled, and $\Ohn_1$ stops rising.
That column, $\Lambda=1/16$, is therefore excluded from the fit. Its
boundary is not closed: at $N=15$ a droplet still escapes at $\Ohn=0.10$, so
that $\Ohn_1\ge0.10$ against $0.113$ from the uncapped law, and whether the law
saturates there or merely bends is not decided by the present data.

\subsection{The shape of a marginal event}

Figure \ref{fig:marginal} shows three events near the boundary and one
thin-shell event, and makes clear
that a single threshold in $\Ohn$ hides several distinct ways in which an
ejection can fail: repeated attempts at end-pinching, of which only the last
succeeds; collisions within a slow and closely spaced train, which decide
individual fates; and a neck that oscillates between thinning and capillary
refilling without ever pinching, so that the tip droplet is reabsorbed. All
place the boundary at the same $\Ohn$ to within $\pm0.005$, which is why the
boundary is sharp although its microscopic physics is not unique. Panel
(\textit{d}) shows the thin-shell regime excluded from the fit: the
punctured shell launches two jets rather than one, the retrograde jet coming
from the backward collapse of the capillary wave on the outer surface.

\subsection{The boundary as a contour of the droplet count}

The boundary is not a bifurcation, and locating it is the one task for which
the droplet count is the appropriate variable. The resolved count falls to
zero continuously, the last non-zero value in every column being exactly one;
the four regimes usually distinguished by inspection (dripping, marginal
ejection, break-up without escape and no break-up) correspond to successive
outcomes of a Poisson variable of that mean; and the contour of unit count agrees with
(\ref{eq:law}) to within a few per cent in every column. The boundary is
therefore the unit contour of a smooth function, and carries its own error
bar, the Poisson variance of a rare event.

\subsection{The distance to the boundary}\label{sec:delta}

We write
\begin{equation}
  \delta=1-\frac{\Ohn}{\Ohn_1(\Lambda)}
  \label{eq:delta-def}
\end{equation}
for the distance to the boundary. It vanishes on it and approaches unity for
inviscid liquids, and it measures how vigorous an event is. The rest of this
paper is written in $\delta$, and it is worth saying why that is possible only
now. An optimum, a scaling or a regime map written in $\delta$ requires the
threshold as a function of the confinement. At a bath the threshold is a
single number, and in the sessile geometry it has not been measured as a
curve \citep{GutierrezEtal2026}. Without (\ref{eq:law}), the results that
follow could be stated only in $\Ohn$, and a statement in $\Ohn$ cannot be
compared between geometries, since each has its own threshold.

\section{What is ejected}\label{sec:ejected}

\subsection{The volume: a law with one constant}\label{sec:mass}

The ejected volume is a single quantity, the liquid that the collapse sets in
motion and throws clear, but it is bounded above in two ways of quite
different nature. The first bound is geometric. At a bath the liquid is
unlimited, yet the mobilized volume is not: what limits it is the cavity,
whose collapse focuses the capillary waves and thereby fixes the scale of the
jet. The gas volume is not a reservoir of ejectable mass but the template that
decides how much liquid the collapse can set in motion, and it does so in
proportion to $V_{\rm gas}$, the energy deposited being fixed by the cavity
(\S\ref{sec:energy}). The proportion implied by classical jet-drop
measurements is of order one per cent: a handful of droplets of radius
$R_d\sim R_0/10$ per event, each of volume $\sim10^{-3}V_{\rm gas}$
\citep{Blanchard1989,Spiel1997}. The second bound is one of mass, and it is
absolute: the event cannot eject more liquid than the drop contains. It is
inactive when liquid is abundant and becomes the only operative constraint as
$\Lambda\to0$, where the same proportion is then taken of $V_{\rm liq}$.

The ejected volume must therefore follow whichever ceiling is lower, and the
simplest smooth expression that reduces to each in turn, without introducing
a second constant, combines them harmonically:
\begin{equation}
  M_e = C\,\delta\;\frac{V_{\rm gas}V_{\rm liq}}{V_{\rm gas}+V_{\rm liq}}
      = C\,\delta\,V_{\rm gas}\left(1-\lambda^{-3}\right),
  \label{eq:redvol}
\end{equation}
where $C$ is a dimensionless constant and the factor $\delta$ expresses the
viscous quenching of the mobilized volume as the boundary is approached. We
write $V_{\rm red}=V_{\rm gas}V_{\rm liq}/(V_{\rm gas}+V_{\rm liq})$ for the
reduced volume, by analogy with two capacities in series. That a single
constant should serve both ceilings is not obvious a priori, and it is the
measurements that establish it; (\ref{eq:redvol}) is, in this sense,
empirical.

\begin{figure*}
  \centering
  \includegraphics[width=\textwidth]{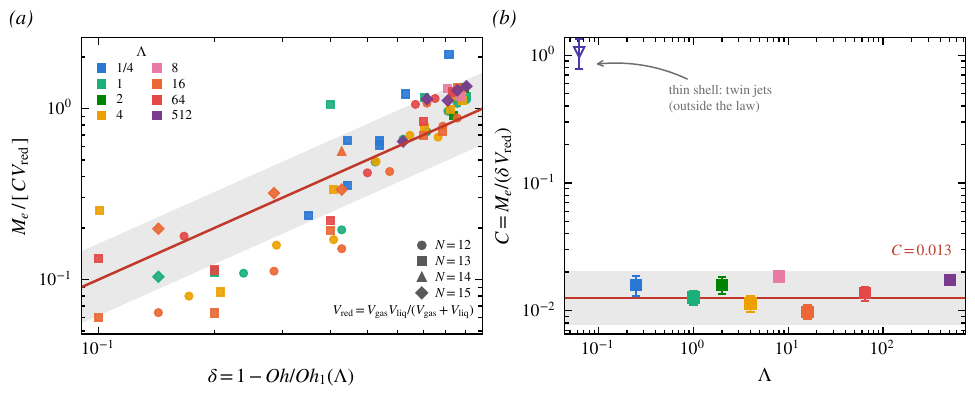}
  \caption{The reduced-volume law (\ref{eq:redvol}). (\textit{a}) Collapse of
  $M_e/(C\,V_{\rm red})$ against $\delta$ for the $71$ physical cases with
  $\Lambda\ge1/4$; colour denotes the column $\Lambda$, symbol shape the level
  $N$. The line is $\delta^1$, the band the residual scatter of a factor
  $1.6$. Points at small $\delta$ lie where the mass is not yet converged in
  mesh (\S\ref{sec:mass}). (\textit{b}) The constant $C=M_e/(\delta V_{\rm
  red})$ column by column, with the standard error of each mean: flat over
  more than three decades of $\Lambda$. The thin-shell column lies a factor
  $85$ above, in the twin-jet regime the law does not describe.}
  \label{fig:mass}
\end{figure*}

Equation (\ref{eq:redvol}) was tested on the $74$ completed ejections with
$\delta\ge0.1$ and $\Lambda\ge1/4$, $71$ physical cases once replicas are set
aside, $43$ of them at $N\ge13$, using escaped
mass only (figure \ref{fig:mass}). With $C$ as the only adjustable parameter,
\begin{equation}
  C=0.013\pm0.005,\qquad \Lambda\ge1/4 ,
  \label{eq:C}
\end{equation}
and the residual scatter is a factor $1.6$. That scatter is not numerical
noise: replicas reproduce the ejected mass to $2\,\%$, and three pairs of runs
of identical physics and different numerical realization differ by factors
$1.06$, $1.83$ and $1.66$. It is the scatter of individual events, the mass
being dominated by the fates of the few largest droplets. The evidence for the
law is the behaviour of the constant it extracts: averaged within each column,
$C=M_e/(\delta V_{\rm red})$ lies between $0.010$ and $0.019$ from
$\Lambda=1/4$ to $512$, with no trend over more than three decades
(figure \ref{fig:mass}(\textit{b})). The alternatives fail on the same test. An
unconstrained power law $M_e/V_{\rm liq}=c_0\Lambda^{a}\delta^{b}$ returns
$a=-0.80\pm0.04$ with a larger residual despite one more parameter and a
prefactor that drifts with $\Lambda$, the effective $a$ being an average of
the local slope $-\Lambda/(1+\Lambda)$ that (\ref{eq:redvol}) implies across
the crossover; and embedding (\ref{eq:redvol}) in the family
$M_e=C\,\delta\,V_{\rm gas}(1+\Lambda^{m})^{1/m}$, $m<0$, which enforces the
same two ceilings with a variable sharpness, a free fit returns $m=-0.99$,
bootstrap interval $[-1.37,-0.77]$.

The exponent of $\delta$ deserves a separate word, because the data appear to
contradict it and do not. Fitting $M_e\propto\delta^{\,b}$ gives $b=1.62$ on
the $N=12$ runs, $1.28$ on $N=13$ and $0.67\pm0.20$ on $N=15$: the exponent
drifts with the mesh and the finest level lies within two standard errors
of unity. The
mechanism is measurable. A coarse mesh loses mass near the boundary, where
the ligament is thinnest, the fine-to-coarse ratio growing as
$\delta^{-0.72}$ (a factor $3.1$ at $\delta=0.14$, against $1.05$ on
average above $\delta=0.6$), and that alone manufactures most of the
apparent excess. The linear quenching of (\ref{eq:redvol}) stands, and the
statement that the ejected mass converges under refinement holds for
$\delta\gtrsim0.4$; closer to the boundary it is a statement about the finest
meshes only.

Expressed as a fraction of the available liquid, (\ref{eq:redvol}) reads
$M_e/V_{\rm liq}=C\,\delta/\lambda^{3}$, which spans four orders of magnitude:
$2.3\times10^{-5}$ at $\Lambda=512$ with $\delta=0.9$, while the measured
fraction exceeds a third at $\Lambda=1/16$. The latter belongs to a regime (\ref{eq:redvol}) does not
describe. Below $\Lambda\simeq1/4$ the shell is thin, the ejection changes
character, and the event ejects $34$ to $40\,\%$ of its liquid almost
independently of $\Ohn$, through the pair of opposed jets of figure
\ref{fig:marginal}(\textit{d}).

\subsection{The largest droplet: a ceiling set by the cavity}\label{sec:rmax}

The geometric ceiling of \S\ref{sec:mass} is visible directly in the largest
droplet of each event. At $\Lambda=16$ and $N=15$, over eight values of $\Ohn$
spanning $\delta=0.14$ to $0.43$, with replicas averaged,
\begin{equation}
  R_{\max}=(0.105\pm0.004)\,R_0 ,
  \label{eq:rmax}
\end{equation}
constant to $3.4\,\%$, while in viscous units the same droplets vary by a
factor $2.4$ (figure \ref{fig:crest}(\textit{c})). The largest droplet scales
with the bubble and not with $\ell_\mu$, and at this column the classical
estimate $R_d\sim R_0/10$ is a measurement. This ceiling is not conservation:
at $\Lambda=16$ the liquid is abundant, and nothing in the available volume
limits the size. It is the cavity, and it bounds the census of
\S\ref{sec:census} from above. Its level, however, depends on the
confinement. At $N=15$ and wherever $\ell_\mu$ is resolved, the median of
$R_{\max}/R_0$ is $0.070$ at $\Lambda=1$, $0.090$ at $\Lambda=4$ and $0.106$ at
$\Lambda=16$, and it moves by $20$ to $30\,\%$ between $N=13$ and $15$ at
$\Lambda=1$ and $16$; at $\Lambda=512$ every escaping run at $N=15$ lies below
the viscous resolution. The ceiling is set by the cavity, and it rises from
thin to thick shells towards the value (\ref{eq:rmax}).

\subsection{The first droplet, and its minimum in $\delta$}\label{sec:crest}

The first droplet (at the tip of the train, in the sense of
\S\ref{sec:census-def}) is the one whose size and speed the scaling laws of
bursting address \citep{GananCalvo2017,Berny2021}, and at a bath its radius
passes through a minimum as $\Ohn$ decreases \citep{GananCalvo2017}. The
minimum survives confinement. At $\Lambda=16$ and $N=15$, with up to four
replicas per point, $R_{d1}/\ell_\mu=10.19$, $9.47$, $9.36$, $9.82$, $10.13$,
$11.20$, $15.04$, $17.74$ and $39.5$ at $\delta=0.14$, $0.15$, $0.20$, $0.25$,
$0.29$, $0.30$, $0.40$, $0.43$ and $0.60$: a minimum at $\delta\simeq0.2$, of
depth $1$ to $5\,\%$ against its neighbours. It is shallow,
and the replicas are what make it measurable at all, their dispersion being
$11$ to $13\,\%$ once the first droplet is identified correctly. Identifying it
as the earliest born instead admits satellites of forked pinch-offs born ahead
of the tip \citep{Berny2022}, and inflates the dispersion between replicas to
a factor six.

Two corrections that could have moved the minimum do not. First, in-flight
coalescence: a quarter of all escaped droplets ($23\,\%$) merge with another
during flight, but measured on the stationary part of its trajectory the
radius of the tip droplet changes by more than $10\,\%$ in only two runs of
that column, and in both the correction deepens the minimum without
displacing it. Second, the time the head of the ligament takes to consolidate
into a droplet (a median of $0.055$ and at most $0.13$ capillary times,
against a flight of about $1.5$) grows with $\delta$, and it is this delay,
not a merger, that made the tip appear to change volume early in its flight at
large $\delta$.

\begin{figure*}
  \centering
  \includegraphics[width=\textwidth]{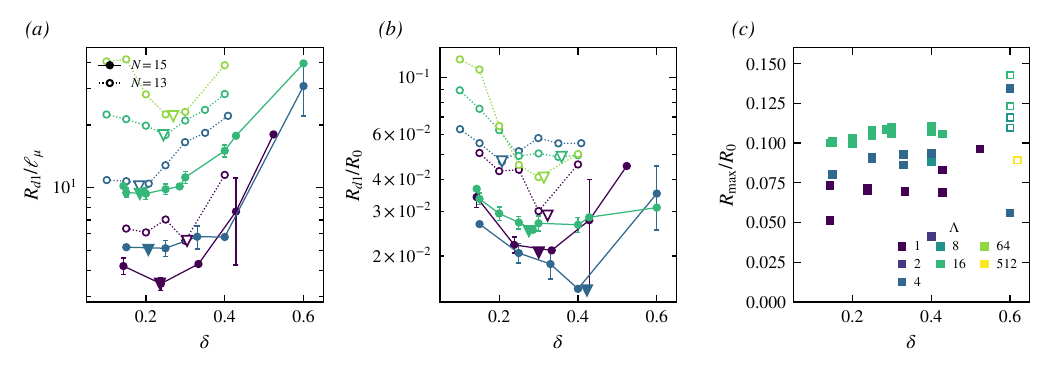}
  \caption{The first and the largest droplet. (\textit{a}) Radius of the first
  droplet in units of $\ell_\mu$ against $\delta$, columns $\Lambda\ge1$,
  $N=15$ filled and joined by solid lines, $N=13$ open and dotted; replicas
  averaged, with their standard error. Triangles mark the minimum of each
  series. (\textit{b}) The same droplets in units of $R_0$: the minimum moves
  to larger $\delta$, as (\ref{eq:units}) requires. (\textit{c}) Largest
  droplet in units of $R_0$ at $N=15$; open symbols, $\ell_\mu<\Delta$.}
  \label{fig:crest}
\end{figure*}

The position of the minimum does not depend on the confinement (figure
\ref{fig:crest}(\textit{a})). Located by a parabola in $\ln R_{d1}$ through
the smallest sampled value and its neighbours, it lies at $\delta=0.236$,
$0.206$ and $0.185$ for $\Lambda=1$, $4$ and $16$ at $N=15$, and at $0.30$,
$0.18$, $0.25$ and $0.27$ for $\Lambda=1$, $4$, $16$ and $64$ at $N=13$.
The mean and spread of the three values at $N=15$ are
\begin{equation}
  \delta^*=0.21\pm0.03 ,
  \label{eq:deltastar}
\end{equation}
constant over a factor sixty-four in $\Lambda$. Confinement does not displace
the optimum; over this range it only flattens it. At the thinnest column,
$\Lambda=1/4$, the minimum falls at the edge of the sampled range and is not
measured.

The minimum sits elsewhere when the same droplets are measured in units of the
bubble (figure \ref{fig:crest}(\textit{b})), and it must. Writing
$F(\delta)=R_{d1}/\ell_\mu$, with $\ell_\mu=\Ohn^2R_0$ and
$\Ohn=\Ohn_1(1-\delta)$,
\begin{equation}
  \frac{R_{d1}}{R_0}=F(\delta)\,\Ohn_1^2\,(1-\delta)^2 ,
  \label{eq:units}
\end{equation}
whose minimum lies where $F'/F=2/(1-\delta)>0$: on the rising branch of $F$,
at larger $\delta$ than $\delta^*$, because the factor $(1-\delta)^2$ keeps
shrinking the droplet in units of $R_0$ after $F$ has begun to grow. The
measured minimum in $R_{d1}/R_0$ lies at $\delta=0.300$ and $0.276$ for
$\Lambda=1$ and $16$ at $N=15$, and at $0.32$, $0.36$ and $0.31$ for
$\Lambda=1$, $16$ and $64$ at $N=13$ (at $\Lambda=4$, sampled more sparsely,
it falls at $0.42$ and $0.21$ at the two levels, and is left out):
\begin{equation}
  \delta^*_R\simeq0.30\pm0.03 .
  \label{eq:deltastarR}
\end{equation}
Neither location is more fundamental; they are one minimum seen in two units.

The distinction decides the comparison with the unbounded bath. There the
optimum is quoted as the Laplace number $La=\Ohn^{-2}$ at which the droplet
radius, at fixed bubble radius, is smallest \citep{Deike2018,GutierrezEtal2026}:
a minimum of $R_d/R_0$. Converted to $\delta$ with a threshold of matching
provenance it is $0.265$ (the optimum $La\simeq1000$ with our threshold
$\Ohn_c=0.043$), $0.293$ (optimum and threshold of \citet{Deike2018}) or
$0.329$ (the optimum $La\simeq1110$ of \citet{GutierrezEtal2026} with the
threshold of \citet{Deike2018}). All three lie within, or at the edge of, the
error of
(\ref{eq:deltastarR}). Measured in the same units, the confined drop places the
optimum where the bath does, and the spread among the literature values is as
large as our error bar.

The size at the minimum is less well defined than its position. Between
$N=13$ and $15$ it falls by a factor $1.6$ to $2$ in every column, and at
$N=15$ it rises with $\Lambda$: $3.5$, $5.0$ and $9.3\,\ell_\mu$
at $\Lambda=1$, $4$ and $16$. These are upper bounds, not converged values. At a
bath the smallest measured first droplet is somewhat above $20\,\ell_\mu$; in the
sessile, confined experiments of \citet{GutierrezEtal2026} it reaches
$7\,\ell_\mu$; the free hollow drop lies at or below the confined value. Those
authors attribute the smoothing of the minimum by confinement to dissipation at
the solid substrate and the pinned contact line. The free drop has neither, and
the minimum is flattened nonetheless, so that explanation cannot be the whole
of it.

For seawater, $\ell_\mu\simeq15.5$\,nm (table \ref{tab:scales}), and
$\delta^*=0.2$ at $\Lambda=16$ is $\Ohn=0.056$, $La=319$: a bubble of about
$5\,\mu$m. The optimum of ejection lies among micrometre-sized bubbles.

\subsection{Two scales, and a ramp between them}\label{sec:ramp}

The first and the largest droplet belong to different scales. $R_{\max}$ is
constant in units of $R_0$ and varies by a factor $2.4$ in units of $\ell_\mu$;
$R_{d1}$ is three to four times smaller and carries the minimum in $\delta$ of
\S\ref{sec:crest}, at the same $\delta$ in every column. The largest droplet is
geometric; the first is set by how far the event lies from the boundary.
Within one column the window in $\Ohn$ is too narrow, a factor $1.4$, for the
dispersion of $R_{d1}$ alone to decide its scale, because the minimum lies
inside it.
Between them the census is not bimodal but a ramp: the radius at birth grows
monotonically with the birth time, from $6$ to $13\,\ell_\mu$ at $t\simeq0.4$--$0.5$
to $25$ to $60\,\ell_\mu$ at $t\simeq0.9$--$1.1$, and the apparent bimodality of
a histogram is that ramp projected on the radius. The fragments too small to
be resolved are late, born at $t=0.83$ to $1.25$: satellites of the final
break-up, not debris of the advancing tip.

\section{The census: no lower cut-off}\label{sec:census}

Between the first droplet and the largest lies the census of every unique
droplet the event produces. Its upper end is the geometric ceiling of
\S\ref{sec:rmax}. Its lower end is the question.

\subsection{Why $\ell_\mu$ cannot be a lower cut-off}\label{sec:why}

The case against the cut-off can be made before any simulation is examined,
and it rests on what $\ell_\mu$ is. Droplet size is set by the local radius
of the ligament at the moment capillarity overcomes whatever stretches it.
The viscocapillary length is a different object: it is the neck radius at
which the pinch-off singularity changes similarity regime, from the
inertial--capillary form of \citet{DayHinchLister1998} to the universal
inertial--viscous solution of \citet{Eggers1993}. One belongs to the parent
thread, the other to the last instants of each individual pinch, and to
identify them is a category error.

Neither terminal regime carries a length of its own. Both are linear in the
time $\tau$ to pinch-off, $h=0.0709\,\sigma\tau/\mu$ in Stokes flow
\citep{Papageorgiou1995} and $h=0.0304\,\sigma\tau/\mu$ in the
inertial--viscous case \citep{Eggers1993,BrennerListerStone1996}, and both run
to $h\to0$. It might be objected that inertia disappears as the neck thins,
leaving a Stokes problem whose scale is $\ell_\mu$. It does not: the neck is
slender, and the Reynolds number that enters the momentum balance carries the
aspect ratio. With the axial scale of the inertial--viscous solution,
$Re_z=\rho vz/\mu=\rho\sigma\ell_\mu/\mu^2=1$ identically, at every $\tau$, so
that inertia sits at order unity however small $h$ becomes, which is the
statement that the regime admits no scale. Which of the two regimes is
terminal has been settled \citep{LiSprittles2016,RothertEtal2003} and does not
matter here, since the approach proceeds through a succession of scale-free
transients \citep{CastrejonPitaEtal2015,LagardeEtal2018}. In three decades of
work on capillary pinch-off
\citep{ListerStone1998,ZhangLister1999,CohenEtal1999,DoshiEtal2003}, no
similarity solution has produced a smallest droplet.

Two mechanisms then carry the cascade below $\ell_\mu$ without leaving the
continuum. Stretching delays Rayleigh--Plateau and lets the ligament thin far
below the radius at which an unstretched thread would break
\citep{EggersVillermaux2008}; and each asymmetric pinch leaves behind a thread
that breaks again, the iterated cascade observed by \citet{ShiBrennerNagel1994}
and \citet{StoneLeal1989}, explained by \citet{BrennerShiNagel1994}, and
reproduced by computation
\citep{EggersDupont1994,WilkesEtal1999,ChenNotzBasaran2002}. None of these
studies reports a smallest satellite; the sequence ends where the optics or
the mesh ends. The floor that does exist is thermal, $\ell_T=(k_BT/\sigma)^{1/2}$
\citep{MoselerLandman2000,ZhaoEtal2020}, at $0.24$\,nm for water against
$\ell_\mu\simeq14$\,nm, so that nearly two decades of deterministic continuum
separate the two. A cut-off calibrated on one liquid is not even transferable
to another: $0.1\,\ell_\mu$ amounts to $6\,\ell_T$ for water but to $260\,\ell_T$
for a $5$\,cSt silicone oil.

\subsection{The floor follows the mesh}\label{sec:mesh}

\begin{figure*}
  \centering
  \includegraphics[width=\textwidth]{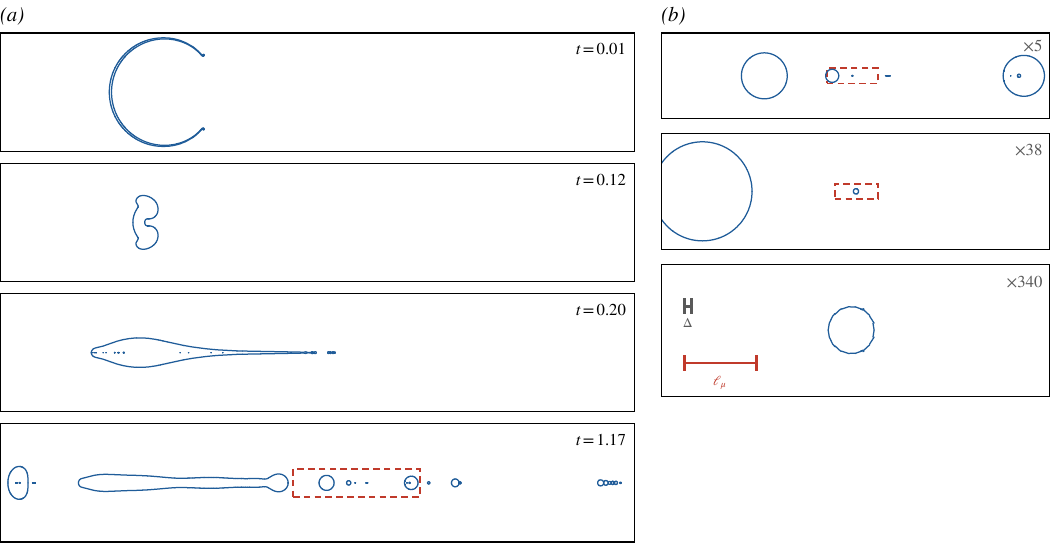}
  \caption{Impulsive capillary jets and the sizes they leave behind
  ($\Ohn=0.08$, $\Lambda=1/16$, $N=15$). (\textit{a}) Four instants in a
  common window: the structure lengthens and sheds fragments, but the range of
  sizes present does not move. (\textit{b}) Three nested magnifications of the
  last instant. The final fragment has $R_d=0.31\,\ell_\mu$ and $3.2\,\Delta$:
  smaller than the viscocapillary length and still spanned by the mesh. Bars
  give $\ell_\mu$ and $\Delta$.}
  \label{fig:scales}
\end{figure*}

\begin{figure*}
  \centering
  \includegraphics[width=\textwidth]{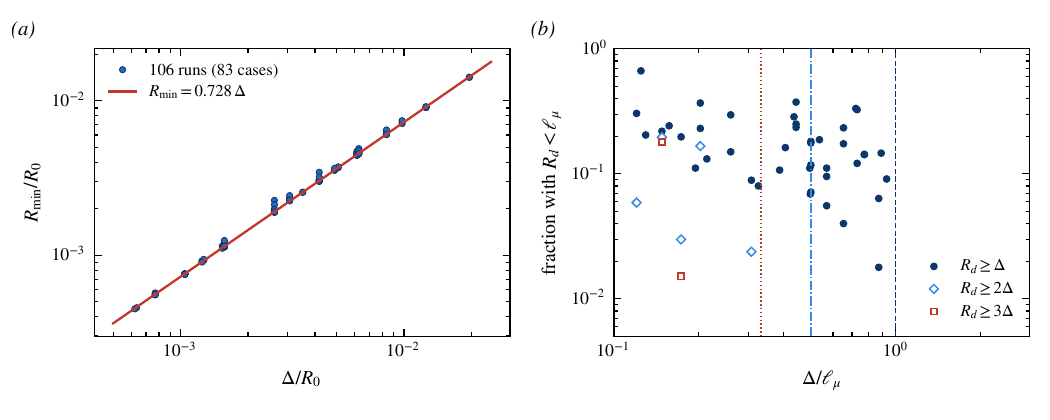}
  \caption{The floor a computation displays is the mesh. (\textit{a})
  Smallest recorded fragment against cell size, over the $83$ cases with at
  least ten resolved droplets; the line is (\ref{eq:rmin}). (\textit{b})
  Fraction of the droplets below $\ell_\mu$ against $\Delta/\ell_\mu$, counting
  only droplets with $R_d\ge\Delta$ (filled circles), $R_d\ge2\Delta$
  (diamonds) and $R_d\ge3\Delta$, the resolved census (squares); the vertical
  lines mark $k\Delta=\ell_\mu$ for each threshold, to the left of which that
  threshold can see below $\ell_\mu$. No run with $\Delta>\ell_\mu$ shows any,
  and runs with none do not appear on the logarithmic axis;
  hardening the threshold empties most runs, leaving three with resolved
  sub-$\ell_\mu$ droplets. All levels are shown on purpose: the content of the
  figure is the dependence on $\Delta$.}
  \label{fig:mesh}
\end{figure*}

Figure \ref{fig:scales} shows what the argument looks like in a computation:
a jet that lengthens by a factor of three and sheds fragments throughout, with
the range of sizes present the same at the end as at the beginning, spanning
three decades below $R_0$. Over the campaign, the smallest recorded fragment
is
\begin{equation}
  R_{\min}=0.729\,\Delta ,
  \label{eq:rmin}
\end{equation}
over the $83$ cases with at least ten resolved droplets, with an interquartile
range of $2.3\,\%$ of the median and not one case outside $[0.5,1.5]\,\Delta$,
across a factor $32$ in $\Delta$ (figure \ref{fig:mesh}(\textit{a})); normalized
by $\ell_\mu$ instead, the same quantity spans three and a half decades. A
volume of $\tfrac43\pi R_{\min}^3\simeq1.6\,\Delta^3$ is the fragment-detection
floor of the interface reconstruction, not a property of the fluid.

The direct test is figure \ref{fig:mesh}(\textit{b}). At the level of the full
census, sub-$\ell_\mu$ droplets appear as a matter of course wherever
$\Delta<\ell_\mu$: over $55$ runs, a median of $34\,\%$ of the population, up to
$64\,\%$. Most of them lie within two cells of the floor. Counting only
droplets with $R_d\ge\Delta$, the median fraction below $\ell_\mu$ is $14\,\%$
over the $49$ runs with $\Delta<\ell_\mu$ that hold at least five such
droplets; with $R_d\ge2\Delta$ it is zero in the median over the $23$ runs
with $2\Delta<\ell_\mu$, though a few runs retain some; with $R_d\ge3\Delta$,
three of the fifteen runs with enough droplets to test do. The sub-$\ell_\mu$ population
is concentrated where a floor set by the mesh puts it, so the conservative
count is the one that matters. Of the completed runs, eighteen satisfy $3\Delta<\ell_\mu$ and can
therefore certify a droplet below $\ell_\mu$, and three contain resolved ones.
Their contributions are unequal in a way that matters. Two are thin-shell runs
at $\Lambda=1/16$ and contribute a single droplet each, at $0.97$ and
$0.38\,\ell_\mu$. The third, at $\Lambda=4$, contributes thirty-six, a fifth
of its resolved census, the smallest at $0.45\,\ell_\mu$ and three cells in
radius: resolved droplets below $\ell_\mu$ are not a curiosity of the thinnest
shells, though that run is a marginal ejection, $\delta=0.006$, in the regime
of repeated failed pinches of figure \ref{fig:marginal}. Each is identifiable
in the trajectory record as the recoil satellite of an asymmetric end-pinch.
That the other fifteen contain none is expected rather than contrary: a census
of ten to twenty droplets need not reach below $\ell_\mu$ merely because the
mesh would allow it. What the test establishes is existence, not frequency.

The count and the mass respond to the mesh in opposite ways, and eleven pairs
of runs at the same $(\Ohn,\Lambda)$ and two levels show it directly. The
ejected mass of the coarse member is a median $0.93$ of that of the fine one,
the outliers lying at the boundary: a coarse mesh adds noise to the mass
without biasing it appreciably. The count, compared at a common physical
cut-off radius, is always larger on the finer mesh, by a median factor $1.30$,
and the deficit of the coarse mesh grows to a factor $5$ to $8$ as the
boundary is approached, where near-marginal pinch-offs are lost when the neck
is under-resolved; in five of nine overlapping pairs with $\delta<0.5$ the
coarse mesh assigns the less ejective regime. Boundaries inferred from coarse
surveys are therefore lower bounds on $\Ohn_1$.

The corollary is uncomfortable for any cut-off inferred from simulation: a
fitted lower cut-off is confounded with the mesh unless it is demonstrably
larger than $\Delta$ over the whole fitted range. The parameter
$x_1=0.1\,\ell_\mu/R_0$ of \citet{GHL2026}, fitted to $N=13$ simulations
\citep{LopezHerreraGanan2024}, lies between $3$ and $60$ times \emph{below} the
cell size across its fitted range: it is not measured but extrapolated. The
direction is right, since droplets do form below the smallest resolved scale,
but nothing anchors the endpoint at a multiple of $\ell_\mu$ rather than at the
thermal length.

A caution that cuts against our own convenience belongs here. The
memorylessness diagnostic used below is itself mesh-dependent, and in the
direction that flatters the exponential on coarse grids: the excess over a
threshold of any light-tailed distribution tends to an exponential once the
threshold is high enough, and a coarse mesh raises the threshold $3\Delta$.
The mean-excess function rises by a median $8\,\%$ when all levels are pooled,
$11\,\%$ for $N\ge13$ and $16\,\%$ for $N\ge14$ (appendix \ref{app:levels}).
The residual does not converge to zero with the levels available; the
statistics below are therefore restricted to $N\ge13$ and quoted with that
drift stated rather than averaged away. Finer meshes than ours would be needed
to close it, and the claim that follows is correspondingly an indication and
not a proof.

\subsection{The census is exponential, with the scale of the cavity}
\label{sec:exp}

It is necessary to say first which distribution is meant, since this
reconciles what would otherwise look like several disagreements. An
instantaneous in-flight population is weighted by residence time and holds
transient fragments; the census of unique droplets, each counted once at its
mature size, is what a spray flux requires. The strongly skewed instantaneous
distributions of \citet{GHL2026} are not in contradiction with what follows:
they are a different statistic. A second distinction matters more.
\citet{Berny2021} describe the jet drops of a bursting bubble as an indexed
sequence (first drop, second, third) and fit the aggregate with a gamma
distribution of order four, a peaked law. Their census enumerates the drops of
the jet, typically between one and fifteen, of which the first is the fastest
and the most robust but not necessarily the largest; ours admits every
fragment that escapes, satellites included, and those fill the small end. The
two are consistent where both are measured, and the difference of shape
follows from what is counted. Two remarks apply. An enumeration presupposes a
largest index and therefore a smallest drop, which is the assumption at issue;
and a census truncated from below at $3\Delta$ acquires a hole that a peaked
family fits naturally, so that the order of the fitted gamma should rise with
$\Delta/\ell_\mu$: a prediction, not a rebuttal.

\begin{figure*}
  \centering
  \includegraphics[width=\textwidth]{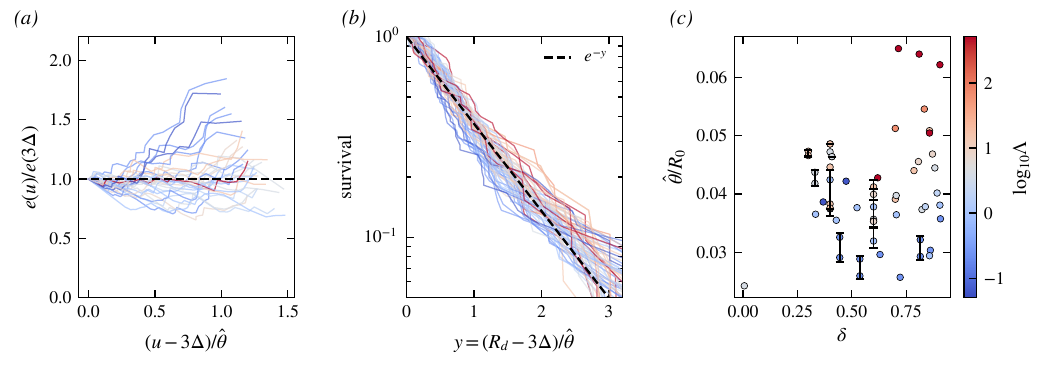}
  \caption{The census is exponential, with a scale set by the cavity.
  (\textit{a}) Mean-excess functions $e(u)=E[R_d-u\,|\,R_d\ge u]$ of the
  $30$ censuses with at least forty resolved droplets at $N\ge13$, divided by
  their value at the truncation, $e(3\Delta)=\hat\theta$, and plotted against
  the threshold $u$ measured from the truncation in units of $\hat\theta$, up
  to the third quartile of each census: a flat curve is an exponential.
  (\textit{b}) Survival functions of the $59$ censuses at $N\ge13$ with at
  least twenty resolved droplets, rescaled by their own $\hat\theta$, against $e^{-y}$, down to $0.05$, below
  which a census retains only one or two droplets. Colour denotes the
  confinement $\Lambda$: the curves that depart upward in (\textit{a}) are the
  thin-shell columns $\Lambda<1$, where a second mechanism adds a coarse tail.
  (\textit{c}) $\hat\theta/R_0$ against $\delta$, with (\ref{eq:theta}); the
  trend with $\delta$ within a column, $\theta\propto(1-a\Ohn)^{-1}$, is smaller
  than the dispersion between cases; error bars are the dispersion among replicas of the same
  physical case.}
  \label{fig:exp}
\end{figure*}

Truncation is treacherous: above a sufficiently high threshold the excesses
of any light-tailed distribution tend to an exponential, so an exponential fit
above one threshold proves little. The diagnostic immune to this is the
mean-excess function $e(u)=E[R_d-u\,|\,R_d\ge u]$, constant if and only if the
distribution is exponential \citep{Embrechts1997} (figure
\ref{fig:exp}(\textit{a})). Applied across the campaign it separates two
families rather than confirming one. Its rise between the truncation and the
third quartile is given column by column in table \ref{tab:rise}.

\begin{table*}
  \begin{center}
  \def~{\hphantom{0}}
  \begin{tabular}{lcccccccc}
  $\Lambda$ & $1/16$ & $1/4$ & $1$ & $2$ & $4$ & $8$ & $16$ & $512$\\[3pt]
  rise (\%) & $60$ & $35$ & $5$ & $-17$ & $-8$ & $15$ & $16$ & $15$\\
  \end{tabular}
  \caption{Rise of the mean-excess function between the truncation and the
  third quartile, median per column $\Lambda$.}
  \label{tab:rise}
  \end{center}
\end{table*}

For $\Lambda\ge1$ the median is $4\,\%$ over $21$ censuses; for $\Lambda<1$ it
is $49\,\%$ over nine, an order of magnitude more, and the departure grows monotonically
as the shell thins, while above unity the columns stay within $\pm20\,\%$ with
no trend. The reason is mechanical. When the liquid is finite and scarce, the
parent drop is itself stretched by the inertia of the radial collapse, and the
ligament it forms carries a scale of its own, adding a coarse tail foreign to
the jet the cavity launched: the twin-jet regime of figure
\ref{fig:marginal}(\textit{d}), which also takes the thin shells outside the
mass law (\S\ref{sec:mass}). A second scale breaks the exponential, which is
what the argument of \S\ref{sec:why} predicts; that it breaks exactly where a
second mechanism is known to appear is the strongest test the argument has
passed. We therefore restrict what follows to $\Lambda\gtrsim1/4$, as the mass
law is restricted, and for the same reason.

For $\Lambda\ge1$ the mean-excess function is flat to within $\pm20\,\%$ over
the lower half of the census in sixteen of the twenty-one censuses, across a
truncation point that varies by a factor $32$. The competing family can be
tested from within, since the generalized inverse Gaussian reduces to a gamma
when its lower cut-off is removed and to an exponential when in addition its
shape exponent is unity: in left-truncated maximum likelihood the lower
cut-off rarely earns its parameter, and by the Akaike information criterion the plain
exponential is preferred in the large majority. Rescaled by their own
$\hat\theta$, the censuses collapse onto $e^{-y}$ over nearly two decades
(figure \ref{fig:exp}(\textit{b})). The fitted Weibull exponent has median
$1.16$, against the value $3$ that volume-weighted Poisson statistics on a
corrugated ligament predict \citep{KooijEtal2024}, and against the peaked
censuses long reported for ligament-mediated fragmentation under sustained
stretching \citep{EggersVillermaux2008}.

The discrepancy is too large to be a technicality, and we read it as causal:
the way a ligament is loaded selects the shape of the census it leaves.
Sustained stretching imprints the corrugation of the parent thread upon its
daughters, and the resulting sizes are correlated and peaked. An impulsive
collapse loads the ligament once, the memory of the parent structure is lost
in the sequence of pinches that follows, and the census that survives is
exponential. That form is not one alternative among several: of all
distributions on the positive axis with a prescribed mean radius, the
exponential is the one of greatest entropy. \citet{Villermaux2007} objects to
maximum-entropy arguments in fragmentation on the grounds that a prescribed
mean \emph{volume} yields $\exp[-(R/\langle R\rangle)^3]$, far too sharp to
describe measured sprays. That objection stands and does not apply here, the
constraint being a mean radius; and we do not derive the exponential and then
look for it, but measure it, and only then observe that it is the form
carrying least information beyond a single scale.

What remains is the scale, and it is set by the cavity. Over the $46$ physical cases with at least twenty resolved droplets,
the fitted scale $\hat\theta=\langle R_d\rangle-3\Delta$ is
\begin{equation}
  \theta\simeq0.039\,R_0 ,
  \label{eq:theta}
\end{equation}
with a residual scatter of a factor $1.25$ (figure \ref{fig:exp}(\textit{c})).
Its dependence on the liquid present is weak: a factor of about two over four
decades in $\Lambda$, against the factor twenty that a scale set by the
available liquid would demand, and it follows the collapse clock of
\S\ref{sec:clock} more closely than any other geometric variable. The scale is
set by the collapsing cavity, as the ceiling of \S\ref{sec:rmax} is, and
$\ell_\mu$ plays no part in it. Within a column it depends on the distance to
the boundary through viscosity alone: fitted with a prefactor for each column
and level over $78$ physical cases, $\theta\propto(1-a\Ohn)^{-1}$, with the free
exponent $-1.27\pm0.21$ and $a=3.8$ the viscous coefficient of the collapse
energy (\S\ref{sec:energy}), while the inviscid energy $K_0(\hat h)$ plays no
part. Viscosity removes a fraction $a\Ohn$ of the energy the collapse
delivers, and the droplets coarsen in the same proportion. The replicas turn that into a measurement: over seventeen
replicated cases the dispersion of $\hat\theta$ within a case is $7.9\,\%$,
against $25\,\%$ between cases, three times smaller, so that $\theta$ is a
property of the case and not of the realization.

The exponential describes the bulk of the census and not its upper end. An
exponential census of $n$ droplets would place its largest near $\theta\ln n$,
which grows by $71\,\%$ from $n=7$ to $n=28$; the measured $R_{\max}$ grows by
$6\,\%$ over the same range, pinned at $2.6\,\theta$ by the ceiling of
\S\ref{sec:rmax}. The upper tail is cut by the cavity, more sharply than an
exponential, and the lower is not cut at all.

The emission of a single event therefore has two ingredients, a volume and a
scale:
\begin{equation}
  n(R;\Ohn,\Lambda)\,\mathrm{d}R =
  \mathcal{N}\,\frac{e^{-R/\theta}}{\theta}\,\mathrm{d}R ,\qquad
  \theta=\frac{\theta_0(\Lambda)}{1-a\Ohn} ,
  \label{eq:joint}
\end{equation}
for $R$ below the ceiling, the number $\mathcal{N}$ being fixed by the volume
(\ref{eq:redvol}) and the scale rather than by a law of its own. Summed droplet
by droplet, the resolved volume goes as $\delta^{1.02\pm0.17}$ at $N=15$, the
linear quenching of (\ref{eq:redvol}); the exponential carries most of it, and
the remainder, which grows with $\delta$ and amounts to a median factor
$1.23$, lies in the part of the census the exponential does not describe. The
volume must therefore be measured through (\ref{eq:redvol}) and never inferred
from a count and a scale; and the absence of a floor does not license an
arbitrarily steep cascade, since surface energy forbids a number density
steeper than $R^{-3}$ as $R\to0$.

\subsection{The axisymmetric artefact, bounded}\label{sec:axi}

An axisymmetric computation forces onto the axis fragments that in three
dimensions would pass one another, and the smaller and faster the fragment,
the less likely that encounter really is. The objection is sound, and it can
be bounded rather than conceded, because the artefact removes population and
never adds any. Every droplet is followed from birth to fate, so the campaign
supplies both limits directly: a ceiling, every droplet ever born, which is
the census obtained if none of these mergers would occur in three dimensions;
and a floor, the same set less those lost to forced coalescence, each droplet
entering at the radius it had when born.

Forced coalescence claims between a sixth and a half of the population, and it
is selective, as the objection asserts: the droplets it takes have a mean
radius $0.55$ of the mean of the ceiling and a coefficient of variation of
$0.41$ against $0.78$, so that removing them raises the mean radius by a
median $17\,\%$. What the selection does not do is change the shape. Since the
two limits are nested they cannot be compared by a two-sample test; the
comparison is instead with chance, by removing at random the same number of
droplets and asking how often the resulting distribution, each normalized by
its own mean, departs from the ceiling by at least as much as the real
thinning does. In the large majority of the censuses with enough droplets to
give the test power, it does not; the exceptions are the cases in which
coalescence claims more than two fifths of the census. The artefact therefore
acts as a rescaling and not as a change of law (which is what an
exponential admits and a peaked distribution would not), and the physical
case, in which fewer of these encounters occur, lies on the ceiling side of
that band. A real breaking wave will hold more free droplets than this
computation, not fewer.

\section{From one event to a spray}\label{sec:spray}

\subsection{The mixture, and its closed form}\label{sec:mixture}

The census of one event is exponential with a scale set by the cavity,
$\theta=cR_0$ with $c\simeq0.039$. The viscous factor of (\ref{eq:joint}) is
set aside here: in a population each bubble carries its own $\Ohn$, and the
factor, between $1$ and $1/(1-a\Ohn_1)$, is absorbed into the mean value of
$c$. A spray is the mixture of such censuses over
the bubble population,
\begin{equation}
  f(R_d)=\int_0^\infty n_b(R_0)\,N(R_0)\,
         \frac{e^{-R_d/cR_0}}{cR_0}\,\mathrm{d}R_0 ,
  \label{eq:mixture}
\end{equation}
with $n_b$ the flux of bursting bubbles and $N$ the number of droplets per
event. Both are measured elsewhere: $N\propto R_0^{\gamma}$ with $\gamma=-1/3$
\citep{Berny2021}, and $n_b\propto\exp[-(R_0/R_s)^{\lambda_b}]$ with
$\lambda_b\simeq1/3$ and $R_s\simeq4.2\,\mu$m, a form that fits breaking-wave
measurements \citep{DeanStok2002} better than a power law over the whole range
\citep{GHL2026}. Nothing in (\ref{eq:mixture}) is adjustable.

The integral has a closed form for every rational $\lambda_b$, and the
rationals are dense, so the restriction is not one. Writing $\lambda_b=n/m$
with $n$ and $m$ coprime positive integers and $a=R_d/cR_s$, the substitution
$R_0=R_su^{m/n}$ followed by a Mellin--Barnes representation produces gamma
functions whose arguments carry the non-integer coefficient $m/n$; rescaling
the Mellin variable by $n$ makes both coefficients integral, and the Gauss
multiplication theorem then absorbs them both, leaving a Meijer $G$ function of
order $n+m$ (appendix \ref{app:mixture}):
\begin{equation}
  f(R_d)\propto
  G^{\,n+m,0}_{\;0,\,n+m}\!\left(\frac{a^{n}}{n^{n}m^{m}}\;\middle|\;
  \begin{matrix}0,\tfrac1n,\dots,\tfrac{n-1}{n},\\[1pt]
  \tfrac{\gamma}{n},\tfrac{\gamma}{n}+\tfrac1m,\dots,
  \tfrac{\gamma}{n}+\tfrac{m-1}{m}\end{matrix}\right).
  \label{eq:meijer}
\end{equation}
The measured case $\lambda_b=1/3$ is $n=1$, $m=3$; the reference case
$\lambda_b=1$ reduces to $f\propto y^{\gamma/2}K_\gamma(2\sqrt{y})$ with
$y=R_d/cR_s$, an exponential mixed over an exponential, which is also why a
generalized inverse Gaussian describes measured spray spectra \citep{GHL2026},
that family being the one this mixture produces.

\subsection{Two limits, and the closure criterion}\label{sec:limits}

At small sizes, independently of $\lambda_b$ and for $\gamma<0$,
\begin{equation}
  f(R_d)\propto R_d^{\,\gamma},
  \label{eq:small}
\end{equation}
a power law all the way down whose exponent is fixed by the number of droplets
per event alone. At large sizes the integrand is sharply peaked at
$R_0^*\propto R_d^{1/(\lambda_b+1)}$ and
\begin{equation}
  f(R_d)\sim\exp\!\left\{-(1+\lambda_b)
  \left[\frac{R_d}{c\lambda_bR_s}\right]^{\lambda_b/(\lambda_b+1)}\right\},
  \label{eq:tail}
\end{equation}
a stretched exponential whose index $\lambda_b/(\lambda_b+1)=n/(n+m)$ is
inherited from the bubbles: $1/2$ for $\lambda_b=1$, $1/4$ for the measured
$1/3$. The bubbles that feed a droplet of radius $R_d$ are those of radius
$R_0^*$, which grows only as $R_d^{1/(\lambda_b+1)}$: the tail of the spray is
fed by an ever narrower band of the bubble population, which is the origin of
the stretched exponential. The algebraic prefactor, needed for any comparison
at small index, is given in appendix \ref{app:mixture}.
Figure \ref{fig:spectrum}(\textit{a}) shows the closed form for four
values of $\lambda_b$, against a direct quadrature of (\ref{eq:mixture}), with
the two limits (\ref{eq:small}) and (\ref{eq:tail}): the smaller the index of
the bubble population, the longer the tail of the spray.

\begin{figure*}
  \centering
  \includegraphics[width=\textwidth]{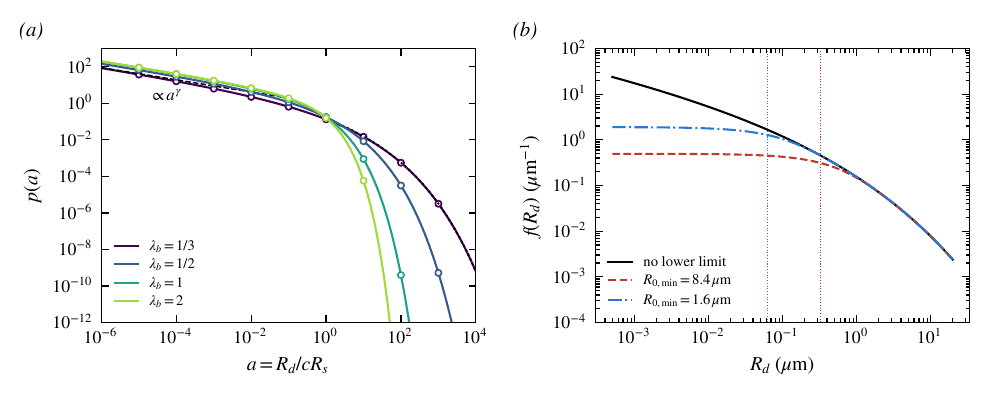}
  \caption{The spectrum of a spray. (\textit{a}) Normalized density
  $p(a)=I(\gamma,a)\,\lambda_b/\Gamma[(1+\gamma)/\lambda_b]$, $a=R_d/cR_s$, for
  $\gamma=-1/3$ and $\lambda_b=1/3$, $1/2$, $1$ and $2$: lines, the closed form
  (\ref{eq:meijer}); circles, direct quadrature of (\ref{eq:mixture}). Dashed
  and dotted, for $\lambda_b=1/3$, the limits (\ref{eq:small}) and
  (\ref{eq:tail}) with the saddle-point prefactor of appendix
  \ref{app:mixture}. (\textit{b}) Seawater, $\lambda_b=1/3$,
  $R_s=4.2\,\mu$m, $c=0.039$: the mixture without a lower limit, and cut at
  the smallest ejecting bubble (\ref{eq:r0min}) for a flat surface
  ($8.4\,\mu$m) and for $\Lambda=1$ ($1.6\,\mu$m); vertical lines at
  $R_d=cR_{0,\min}$.}
  \label{fig:spectrum}
\end{figure*}

Equation (\ref{eq:small}) states the closure criterion exactly. For
$\lambda_b=1$ the normalization is proportional to $\Gamma(1+\gamma)$ and
$\langle R_d\rangle=(1+\gamma)cR_s$, so the emitted \emph{number} converges if
and only if $\gamma>-1$, while the emitted \emph{mass} converges for any
$\gamma>-4$ and surface energy already forbids $\gamma\le-3$. Within
$-3<\gamma\le-1$ the number diverges while mass and energy remain finite: the
asymmetry of \S\ref{sec:mesh} between count and mass, now a condition on a
measurable exponent. The oceanic value lies on the convergent side.

The same mixture with a power-law population returns a power law:
\citet{Berny2021}, with $n_b\propto R_0^{-3/2}$ and a gamma census of order four,
obtain $f\propto R_d^{-5/3}$. The two results differ in kind. A power-law
population gives a scale-free spray; a generalized exponential gives
(\ref{eq:meijer}), a power law crossing over to (\ref{eq:tail}) at
$R_d\sim cR_s\simeq0.2\,\mu$m, within the range where laboratory and field
spectra are reported. The shape of the aerosol spectrum is therefore a
measurement of the bubble size distribution, and the crossover locates
$cR_s$ without a fit.

\begin{figure}
  \centering
  \includegraphics[width=\columnwidth]{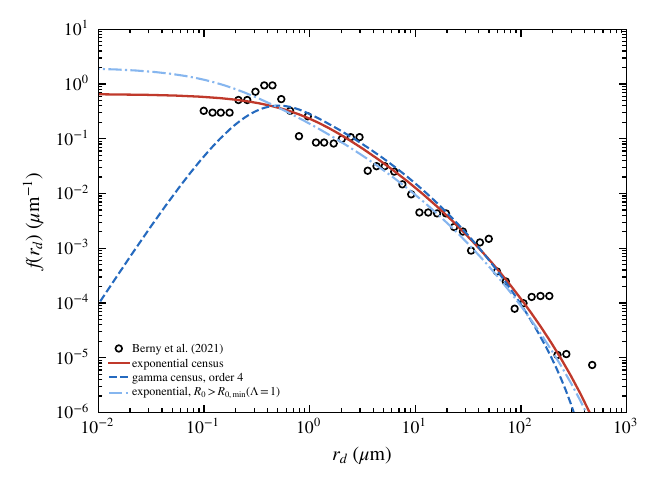}
  \caption{Probability density of the droplet radius. Lines: the mixture
  (\ref{eq:mixture}) with the population of \S\ref{sec:mixture},
  $N\propto R_0^{-1/3}$ and bubbles between $R_{0,c}$ and $5$\,mm, each
  normalized to unit integral: exponential census $\theta=cR_0$ (solid), gamma
  census of order four with the same mean (dashed), and exponential census from
  the confined threshold at $\Lambda=1$ (dash-dotted); seawater at
  $15\,^\circ$C. Circles: the spray spectrum of \citet{Berny2021}, with one
  amplitude fitted to the solid line.}
  \label{fig:berny}
\end{figure}

The comparison can be made directly (figure \ref{fig:berny}). With the
population of \S\ref{sec:mixture}, $N\propto R_0^{-1/3}$, and bubbles from the
flat-surface threshold $R_{0,c}$ (table \ref{tab:scales}) up to $5$\,mm, the
mixture of exponential censuses reproduces the spectrum that \citet{Berny2021}
obtain by combining their simulated per-event statistics with an oceanic
bubble population to within a factor of two (an rms deviation of $0.68$ in
$\ln f$ from $0.1$ to $500\,\mu$m, with one free amplitude). It recovers the
plateau below half a micrometre, which is the image of the smallest ejecting
bubble, and a local slope of $-1.6$ between $1$ and $100\,\mu$m, against the
$-5/3$ of their fit. A gamma census of order four with the same mean, their
own choice, does worse ($1.10$), because it empties the small sizes; moving the
lower limit to the confined threshold at $\Lambda=1$ leaves the spectrum above
a micrometre unchanged and raises the plateau ($0.79$).

What (\ref{eq:mixture}) gives is the spectrum \emph{produced} at the surface,
every droplet counted once at its birth, and it is in that sense that it is
compared with the spectrum of \citet{Berny2021}, obtained in the same way. A
measured aerosol spectrum is a different quantity
\citep{deLeeuwEtal2011,GananCalvo2023,GHL2026,LeibenspergerEtal2026}: it is
what remains of the produced spectrum after the droplets have escaped the
layer next to the surface, been transported, dried and sampled, and it mixes
jet and film drops from a bubble population that is rarely known. The census
includes large and slow droplets born late in the ramp of \S\ref{sec:ramp},
most of which are unlikely to escape. Relating the two spectra requires
weighting each droplet by its probability of escape, which we leave for
future work.

\subsection{The bubbles that the boundary admits}\label{sec:admits}

For a fixed liquid, a threshold in $\Ohn$ is a bubble size, since
$\Ohn=(\ell_\mu/R_0)^{1/2}$. With the boundary (\ref{eq:law}), the smallest
bubble that ejects is
\begin{equation}
  R_{0,\min}(\Lambda)=\frac{\ell_\mu}{\Ohn_1^{2}}
  =\frac{\ell_\mu}{\Ohn_c^{2}}\left(1+\frac{2\beta}{\lambda}\right)^{-2}:
  \label{eq:r0min}
\end{equation}
$8.4\,\mu$m for seawater at a flat surface, $1.6\,\mu$m at $\Lambda=1$, a factor
of five in radius and about $150$ in volume. In a breaking wave, where the
bursting bubble is normally backed by a finite parcel of liquid rather than by
a half-space, (\ref{eq:r0min}) moves the lower limit of the integration over
the bubble spectrum by a factor of five into its most populated part, a range no
flat-surface experiment enters. The same range holds the optimum of
\S\ref{sec:crest}, near $5\,\mu$m, and a census with no floor places a fifth to
two fifths of the emitted number below the reach of current transport-loss
budgets \citep{WangEtal2026}.

Figure \ref{fig:spectrum}(\textit{b}) shows what that limit does to the
spectrum. Cut at $R_{0,\min}$, the mixture (\ref{eq:mixture}) keeps the power
law (\ref{eq:small}) only down to $R_d\simeq cR_{0,\min}$ and flattens below
it: at $0.33\,\mu$m for a flat surface, at $0.06\,\mu$m for $\Lambda=1$, where
the spectrum holds three to four times as many droplets at $0.01\,\mu$m. The
submicrometre end of the jet-drop spectrum is thus set by the ejection
boundary, and confinement extends it by a factor of five in droplet size. The
cut is idealized: $c$ and $N\propto R_0^{\gamma}$ are taken constant down to
$R_{0,\min}$, whereas the ejected volume vanishes there as $\delta$
(\S\ref{sec:mass}), which softens and lowers the edge. The panel gives the
order of the effect, not the spectrum.

\subsection{Jet and film drops in the submicrometre range}\label{sec:jetfilm}

The disputed share of jet drops in the submicrometre marine aerosol
\citep{WangEtal2017,JiangEtal2022,QuinnEtal2017,deLeeuwEtal2011} turns on the
lower end of the per-event census. Mechanistic generation functions assemble
the spray from jet drops and from film drops in two sub-processes, each
dressed with a per-event gamma distribution of order $m_\Gamma$
\citep{DeikeReichlPaulot2022}, and attribute the submicrometre aerosol almost
entirely to film drops. That attribution depends on one choice that has not
been measured. A gamma density of order $m_\Gamma$ vanishes as $x^{m_\Gamma-1}$
at small size; the value adopted for jet drops, $m_\Gamma=11$, suppresses the
small tail as $x^{10}$, whereas the census measured here is exponential,
$m_\Gamma=1$, and does not suppress it at all. With the parametrization of
\citet{DeikeReichlPaulot2022} left otherwise unchanged, the exponential census
makes jet drops overtake film drops below $r_d\simeq0.1$--$0.2\,\mu$m,
depending on the lower limit of the bubble population, and the jet
contribution at $0.1\,\mu$m changes by twelve orders of magnitude with the
order of the assumed law alone. The same
census can be confronted with a measurement: counts of submicrometre aerosol from
controlled bubbles obtained behind an impactor \citep{JiangEtal2022} fall
below the part of the jet census that such an impactor transmits, by factors
of three to seven (appendix \ref{app:jetfilm}). Between $0.3$ and $3\,\mu$m the
film modes still dominate in that parametrization; we make no claim there.

\section{Discussion}\label{sec:discussion}

\subsection{What the computation does not settle}

Five limitations run through the results, and each is stated where it
applies; we collect them here. The drift of the mean-excess diagnostic with
the level has not converged (\S\ref{sec:mesh}, appendix \ref{app:levels}), so
that the exponential census is an indication supported by a consistent body of
evidence, not a demonstration. The axisymmetric artefact is bounded but not
removed (\S\ref{sec:axi}). The thin shells, $\Lambda<1/4$, are outside the mass
law and the exponential alike, through a second mechanism we identify but do
not model. The puncture radius, tied to the mesh through (\ref{eq:puncture}),
lies at or beyond the range over which insensitivity was verified in the two
thinnest columns, where the mouth becomes larger than the wall; this is why
the thin-shell prefactor of the collapse clock is not established
(\S\ref{sec:clock}), and it bears on figures \ref{fig:scales} and
\ref{fig:hollow}, both drawn at $\Lambda=1/16$. Finally, levels beyond $N=15$
fail at initialization in the thinnest columns and are in any case beyond the
computing power available, so that the question of convergence of the census
is left open rather than closed.

\subsection{Clean liquid}

One assumption underlies all of this: the liquid is clean. It enters twice, in
the rupture thickness set by the van der Waals part of the disjoining pressure,
and in the uniform surface tension carried by the interfaces throughout the
collapse. Neither is guaranteed in the sea, but the balance is not obvious.
Surfactants alter the rupture thickness, and Marangoni stresses oppose the
very deformations that drive the process, resisting the retraction of the
rim, damping the capillary waves before they focus and stiffening the ligament
against pinch-off. Against this, a vigorous sea creates interface at an
enormous rate, diluting a finite amount of surface-active material over an
ever growing area; and at the scales that matter for the fine aerosol, that
surface is renewed faster than a dissolved surfactant can populate it, the
capillary time of a $10\,\mu$m bubble being a few microseconds against the
milliseconds that diffusion requires. Below some size, bubbles in an energetic
sea may therefore burst from surfaces that are effectively clean. The clean
case is the necessary reference, and the contaminated case must be built on it
one effect at a time.

\subsection{The experiment that is missing}

Every prediction of this work that bears on the sea lies in one range of
bubble size, from about one to a hundred micrometres: the smallest bubbles that
confinement admits (\S\ref{sec:admits}), the optimum of the first droplet
(\S\ref{sec:crest}), and the submicrometre droplets where jet and film compete
(\S\ref{sec:jetfilm}). No flat-surface experiment enters it, and the
mechanistic generation functions cut their bubble population above it. The
decisive measurement is a count of the submicrometre aerosol from bubbles of
$La=10^3$ to $10^4$ (radii of about $15$ to $150\,\mu$m in seawater) without an impactor in the line, where the jet census
predicted here and the film attribution in current use differ by the factors
of \S\ref{sec:jetfilm}.

\subsection{The hollow droplet as a product}\label{sec:product}

\begin{figure*}
  \centering
  \includegraphics[width=\textwidth]{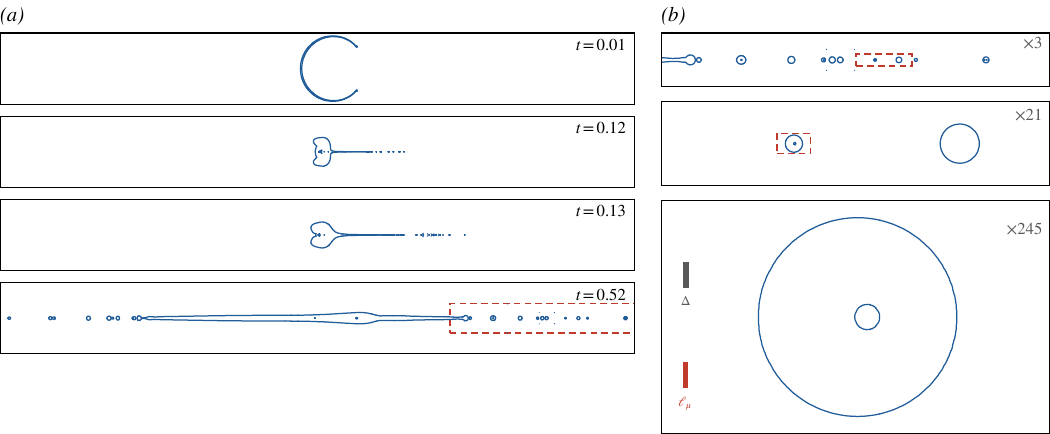}
  \caption{The bursting reproduces its own initial condition ($\Ohn=0.02$,
  $\Lambda=1/16$, $N=15$). (\textit{a}) Four instants: the punctured shell,
  its collapse, the first ligament, and the two opposed jets that a thin shell
  produces. (\textit{b}) Three nested magnifications of the last, ending on a
  hollow droplet, a daughter cavity enclosed by a daughter shell
  ($\Lambda'\simeq540$). The droplet is $65\,\Delta$ across and its cavity
  $8\,\Delta$, so the object is resolved and not a numerical inclusion.
  Here $\ell_\mu=0.64\,\Delta$, below the cell size, so this case carries no
  information about the viscocapillary length.}
  \label{fig:hollow}
\end{figure*}

The course of the event closes on itself. The same break-up that produces the
census also produces hollow droplets (figure \ref{fig:hollow}): a daughter
cavity enclosed by a daughter shell, the configuration of \S\ref{sec:config}
reproduced at smaller scale. Eight are resolved at $t=0.5$ in the case shown,
with $\Lambda'$ between $4\times10^{2}$ and $4\times10^{3}$: solid droplets
carrying a small daughter bubble. Each is a fresh generator, carrying
$\Ohn'=\Ohn\,(R_0/R_0')^{1/2}$, six to twenty times the parent value, so that
here every daughter falls beyond its own boundary and the sequence stops after
one generation. That is a limitation of the parent and not of the mechanism:
since $\Ohn=(\ell_\mu/R_0)^{1/2}$, a parent an order of magnitude larger
(where most of the oceanic bubble spectrum lies) starts far enough below the
boundary to carry some of its daughters at least one generation further, each
adding its own census to the spray.

Only double tangency produces a hollow droplet in the full sense, a daughter
shell around a cavity of comparable size, and we have measured it with the gas
labelled exactly and the mesh refined from $N=13$ to $14$. At $\Ohn=0.04$,
$\Lambda=1/8$ and $\Delta t=0.2\,t_{\rm per}$, the event ejects a droplet of
radius $0.121\,R_0$ enclosing a bubble of $0.052\,R_0$, $43\,\%$ of its radius:
a hollow droplet of its own $\Lambda'=11.7$, leaving at $2.1\,V_2$. Every
quantity agrees between the two levels to within a few tenths of a per cent. Its
own boundary, from (\ref{eq:law}) at $\lambda'=2.33$, is $\Ohn_1'=0.074$,
against its own $\Ohn'=0.18$: $\delta'=-1.4$, beyond the boundary, as the
argument above requires for a parent of this size. A parent at $\Ohn\simeq0.017$
would place it on the boundary.

\section{Conclusions}\label{sec:concl}

A bubble bursting from a hollow droplet ejects wherever $\Ohn$ lies below
$\Ohn_c(1+2\beta/\lambda)$, with $\beta\simeq0.83$: confinement admits bubbles
that a flat surface excludes, through an added Laplace overpressure and a
reduced inertia of the shell, both of order $1/\lambda$. The collapse behind
that boundary is that of a thread of gas pinching next to its mouth, with the
exponent $1/2$ of inviscid thread collapse whether the shell has one mouth or
two, and a tip that obeys $v^2R=\text{const}$ throughout its flight. The
collapse time follows a single saturating law in the shell thickness, which
predicts the columns not used to fit it, and the energy the collapse deposits
is of order $\sigma R_0^2$, set by the cavity to within a factor of two.
Everything the event delivers is set by the cavity or by the distance $\delta$
to the boundary: a volume $M_e=C\,\delta\,V_{\rm red}$ with a single constant;
a first droplet whose minimum lies at $\delta\simeq0.2$ in viscocapillary units
and $\delta\simeq0.3$ in units of the bubble, whatever the confinement, the
latter where the unbounded bath places its optimum; a largest droplet that scales with the bubble,
reaching a tenth of its radius in thick shells; and between
them an exponential census with a scale of four hundredths of the bubble
radius, bounded above by the cavity and not bounded below at all.

The absence of a lower bound is general. An impulsive capillary jet carries no
length of its own; the viscocapillary length is a crossover of the pinch
singularity, and the floor a computation displays is its mesh. Number-weighted
quantities are therefore lower bounds at a declared cut-off, whereas the mass
converges, and the criterion separating them is a condition on a single
measurable exponent. Mixed over a bubble population, the census gives the
spectrum of a spray in closed form for any rational exponent of that
population, and its shape is a diagnostic of the bubbles rather than of the
fragmentation that shaped each event. The course of the event, finally, closes
on itself: the bursting of a hollow droplet produces hollow droplets.

\section*{Acknowledgements}
The photograph of figure
\ref{fig:context}(\textit{a}) was taken by Juan Luis Varela, who generously made
it available for this work. Jos\'e M. L\'opez-Herrera and Miguel A. Herrada
provided highly appreciated computational support. The author is grateful to
David Fern\'andez-Rivas for valuable discussions.

\section*{Use of AI tools}
The author used a large language model (Claude, Anthropic) as an assistant
for post-processing the simulation output, for statistical fitting, for
preparing figures and for editing the text of the manuscript.
The study was designed and conducted by the author, who verified all results and takes full
responsibility for the content.

\section*{Funding}
This work was supported by grant PID2022-140951OB-C21, funded by
MCIN/\allowbreak AEI/\allowbreak 10.13039/\allowbreak 501100011033 and by ERDF/EU.

\section*{Declaration of interests}
The author reports no conflict of interest.

\section*{Data availability statement}
The data that support the findings of this study are available from the author
upon reasonable request.

\appendix

\section{Closed form of the mixture over the bubble population}
\label{app:mixture}

We derive (\ref{eq:meijer})--(\ref{eq:tail}) of \S\ref{sec:spray}. The starting point is the mixture
\begin{equation}
  f(R_d)=\int_0^\infty n_b(R_0)\,N(R_0)\,
         \frac{e^{-R_d/cR_0}}{cR_0}\,\mathrm{d}R_0 ,
  \label{sm:mixture}
\end{equation}
with $N(R_0)=N_0(R_0/R_s)^{\gamma}$ the number of droplets per event and
$n_b(R_0)\propto\exp[-(R_0/R_s)^{\lambda_b}]$ the flux of bursting bubbles.
Throughout we write $\lambda_b=n/m$, with $n$ and $m$ coprime positive
integers, and
\begin{equation}
  a=\frac{R_d}{cR_s},
\end{equation}
so that all lengths are referred to $cR_s$, the only scale the mixture
possesses.

\subsection{Reduction to a Mellin--Barnes integral}

Up to constants, (\ref{sm:mixture}) is
\begin{equation}
  I(\gamma,a)=\int_0^\infty x^{\gamma-1}
  \exp\!\left[-x^{\,n/m}-\frac{a}{x}\right]\mathrm{d}x ,
  \label{sm:u}
\end{equation}
with $\lambda_b=n/m$ and $n$, $m$ coprime positive integers. No generality is
lost by taking $\lambda_b$ rational: the rationals are dense, and a measured
$\lambda_b$ carries a finite uncertainty in any case.

The substitution $x=u^{m/n}$ removes the fractional power from the first
exponential and gives
\[
  I=\frac{m}{n}\int_0^\infty u^{\,m\gamma/n-1}e^{-u-au^{-m/n}}\,\mathrm{d}u .
\]
Inserting
$e^{-w}=\frac{1}{2\pi i}\int\Gamma(t)w^{-t}\mathrm{d}t$ with $w=au^{-m/n}$ and
doing the $u$ integral first,
\begin{equation}
  I=\frac{m/n}{2\pi i}\int\Gamma(t)\,
    \Gamma\!\left(\tfrac{m}{n}(t+\gamma)\right)a^{-t}\,\mathrm{d}t .
  \label{sm:mb}
\end{equation}

\subsection{Removing the coefficients: two applications of Gauss}

The coefficient $m/n$ multiplying $t$ inside the second gamma is what would
make (\ref{sm:mb}) a Fox $H$ function rather than a Meijer $G$, which requires
every gamma to carry unit coefficient in the Mellin variable. When $n=1$ that
coefficient is an integer and one application of the Gauss multiplication
theorem,
\begin{equation}
  \Gamma(pz)=(2\pi)^{(1-p)/2}\,p^{\,pz-1/2}
             \prod_{k=0}^{p-1}\Gamma\!\left(z+\frac{k}{p}\right),
  \label{sm:gauss}
\end{equation}
suffices. For general $n$ it does not, and the step that resolves it is to
rescale the Mellin variable itself: setting $t=ns$ turns the two arguments into
$\Gamma(ns)$ and $\Gamma(m(s+\gamma/n))$, \emph{both} with integer
coefficients, so that (\ref{sm:gauss}) applies to each in turn. Collecting the
constants,
\begin{widetext}
\begin{equation}
  \boxed{\;I(\gamma,a)=\frac{m^{\,m\gamma/n+1/2}}
  {\sqrt{n}\,(2\pi)^{(n+m-2)/2}}\;
  G^{\,n+m,0}_{\;0,\,n+m}\!\left(\frac{a^{n}}{n^{n}m^{m}}\;\middle|\;
  \begin{matrix}0,\tfrac1n,\dots,\tfrac{n-1}{n},\\[1pt]
  \tfrac{\gamma}{n},\tfrac{\gamma}{n}+\tfrac1m,\dots,
  \tfrac{\gamma}{n}+\tfrac{m-1}{m}\end{matrix}\right)\;}
  \label{sm:meijer}
\end{equation}
\end{widetext}
This is a two-integer family of order $n+m$, the first $n$ parameters the
progression $j/n$ and the remaining $m$ the progression $\gamma/n+k/m$. The two
members that matter here are
\begin{align}
  n{=}m{=}1:&\quad G^{2,0}_{0,2}\!\left(a\,\middle|\,0,\gamma\right)
       =2\,a^{\gamma/2}K_\gamma\!\left(2\sqrt{a}\right),
       \label{sm:bessel}\\[2pt]
  n{=}1,\,m{=}3:&\quad \frac{3^{\,3\gamma+1/2}}{2\pi}\,
       G^{4,0}_{0,4}\!\left(\frac{a}{27}\,\middle|\,
       0,\gamma,\gamma+\tfrac13,\gamma+\tfrac23\right),
       \label{sm:n3}
\end{align}
the first the mixture of an exponential over an exponential, the second the
measured case $\lambda_b=1/3$. With $\gamma=-1/3$ the parameters of
(\ref{sm:n3}) are $\{0,-\tfrac13,0,\tfrac13\}$; the repeated entry places the
function in its logarithmic case, which is regular but does not separate into
elementary generalized hypergeometric functions.

Equation (\ref{sm:meijer}) has been checked against numerical quadrature of
(\ref{sm:u}) at forty significant digits for twelve combinations spanning
$n\le4$, $m\le7$, $-1.5\le\gamma\le2$ and $a$ from $5\times10^{-2}$ to $12$,
including cases with $n>1$ and with $\gamma<-1$. It is also consistent under
reduction of the fraction ($n/m=2/4$, $3/9$ and $4/6$ reproduce $1/2$, $1/3$
and $2/3$ to thirty-one digits), so that the coprime form merely gives the
representation of least order.

\subsection{Small sizes}

As $a\to0$ with $\gamma<0$ the integral (\ref{sm:u}) is controlled by $x$ of
order $a$, where $\exp(-x^{n/m})\to1$. Setting $x=ay$ leaves a pure number,
$\int_0^\infty y^{\gamma-1}e^{-1/y}\mathrm{d}y=\Gamma(-\gamma)$, so that
$I\simeq\Gamma(-\gamma)\,a^{\gamma}$ and
\begin{equation}
  f(R_d)\propto R_d^{\,\gamma}\qquad(R_d\ll cR_s,\ \gamma<0),
  \label{sm:small}
\end{equation}
independently of $n$ and $m$; for $\gamma>0$ the density tends instead to a
constant. The small-size exponent is that of the number of
droplets per event and carries no memory of the shape of the bubble
population. This is the limit in which a generalized exponential and a power
law population are indistinguishable, as noted by \citet{GHL2026}.

\subsection{Large sizes, and which bubbles feed the tail}

For $a\gg1$ the exponent of (\ref{sm:mixture}),
$g(R_0)=(R_0/R_s)^{\lambda_b}+R_d/cR_0$, is sharply peaked. Its minimum lies at
\begin{equation}
  R_0^{*}=\left[\frac{R_dR_s^{\lambda_b}}{c\lambda_b}\right]^{1/(\lambda_b+1)},
  \label{sm:saddle}
\end{equation}
and gives $g(R_0^{*})=(1+\lambda_b)\left[R_d/c\lambda_b R_s\right]
^{\lambda_b/(\lambda_b+1)}$, so that
\begin{equation}
  f(R_d)\sim\exp\!\left\{-(1+\lambda_b)
  \left[\frac{R_d}{c\lambda_b R_s}\right]^{\lambda_b/(\lambda_b+1)}\right\}.
  \label{sm:tail}
\end{equation}
The index of the stretched exponential, $\lambda_b/(\lambda_b+1)=n/(n+m)$, is
$1/2$ for $\lambda_b=1$ and $1/4$ for $\lambda_b=1/3$. For $\lambda_b=1$ equation
(\ref{sm:tail}) reduces to $\exp(-2\sqrt{a})$, the large-argument asymptotics
of (\ref{sm:bessel}), as it must.

The algebraic prefactor is worth recording, because at small index the
exponential grows so slowly that the prefactor remains a sizeable fraction of
$\ln f$ over any accessible range. With
$g''(x_*)=\lambda_b(\lambda_b+1)x_*^{\,\lambda_b-2}$,
\begin{widetext}
\begin{equation}
  I(\gamma,a)\simeq\sqrt{\frac{2\pi}{\lambda_b(\lambda_b+1)}}\;
  x_*^{\,\gamma-\lambda_b/2}\,
  e^{-(1+\lambda_b)(a/\lambda_b)^{\lambda_b/(\lambda_b+1)}},
  \qquad x_*=\left(\frac{a}{\lambda_b}\right)^{\!1/(\lambda_b+1)},
  \label{sm:silla}
\end{equation}
\end{widetext}
which approaches (\ref{sm:meijer}) to better than a per cent by $a\sim10^{3}$
for $\lambda_b=2/3$ and $3/4$, and to $0.1\,\%$ by $a\sim10^{7}$ for
$\lambda_b=1/3$. Comparing only the exponential factor, as is tempting, leaves a
discrepancy of some $25\,\%$ in $\ln f$ at $\lambda_b=1/3$ that does not
disappear over any range that can be plotted: the correction is $O(\log a)$
against a leading term growing as $a^{1/4}$.

Equation (\ref{sm:saddle}) carries the physical content. The bubbles that
dominate the emission of a droplet of radius $R_d$ are those of radius
$R_0^{*}$, which grows only as $R_d^{1/(\lambda_b+1)}$: reaching a droplet ten
times larger calls on bubbles only $10^{1/(\lambda_b+1)}$ times larger, of which
the exponential population holds exponentially few. The tail of the spray is
therefore fed by an ever narrower band of the bubble population, and that is
the origin of the stretched exponential.

\subsection{Normalization and moments for $\lambda_b=1$}

For $n=m=1$ the moments follow in closed form from
$\int_0^\infty u^{\mu}K_\nu(u)\,\mathrm{d}u=
2^{\mu-1}\Gamma\!\big(\tfrac{1+\mu+\nu}{2}\big)
\Gamma\!\big(\tfrac{1+\mu-\nu}{2}\big)$ \citep{GradshteynRyzhik}. Writing
$y=R_d/cR_s$ and $p(y)\propto y^{\gamma/2}K_\gamma(2\sqrt{y})$,
\begin{equation}
  \int_0^\infty p\,\mathrm{d}y=\tfrac12\Gamma(1+\gamma),\qquad
  \langle y\rangle=1+\gamma ,
\end{equation}
so the normalized density is
$p(y)=[2/\Gamma(1+\gamma)]\,y^{\gamma/2}K_\gamma(2\sqrt{y})$ and
$\langle R_d\rangle=(1+\gamma)\,cR_s$. The first of these exists only for
$\gamma>-1$, which is the convergence criterion of \S\ref{sec:limits}; the
same calculation with the weight $y^{3}$ shows that the mass converges for any
$\gamma>-4$.

\subsection{Relation to the generalized inverse Gaussian}

For $n=m=1$ the integrand of (\ref{sm:mixture}),
$R_0^{\gamma-1}\exp[-R_0/R_s-(R_d/c)/R_0]$, is a generalized inverse Gaussian
density in $R_0$, whose normalizing constant is precisely
$2(R_dR_s/c)^{\gamma/2}K_\gamma(2\sqrt{a})$. The appearance of a Bessel
function in (\ref{sm:bessel}) is thus not an accident of the algebra but the
statement that the spray spectrum \emph{is} the normalization of a generalized
inverse Gaussian whose parameters slide with $R_d$. Spray spectra have been
described empirically by generalized inverse Gaussian
distributions~\citep{GHL2026}; the present derivation identifies the mixture
that produces that family.

\section{The replica ensemble}
\label{app:replicas}

A replica is the same physical case (identical binary, identical arguments,
identical mesh criteria) executed with a different number of parallel
subdomains. The code is deterministic: repeating a run with the same
decomposition reproduces it bit for bit. Changing the decomposition changes the
order in which floating-point reductions are performed, and therefore the last
bit; the pinch-off amplifies that difference into a divergent trajectory. The
resulting spread is not a numerical error bar but an ensemble over microscopic
perturbations, whose physical analogue is real: two nominally identical bubbles
differ by vibration, thermal noise and initial shape. An equivalent ensemble
built on an explicit $1\,\%$ noise in the initial velocity field was introduced
for bubble bursting by \citet{Berny2022}, who verified
insensitivity to the amplitude and the type of the perturbation.

Seventeen physical cases were replicated, with two to four realizations each.
The dispersion is strongly dependent on the quantity (table \ref{tab:replicas}).

\begin{table*}
  \begin{center}
  \def~{\hphantom{0}}
  \begin{tabular}{lc}
  Quantity & Dispersion within a case\\[3pt]
  ejected mass & $0.6$--$1.8\,\%$\\
  volume-weighted mean radius & $2$--$3\,\%$\\
  arithmetic mean radius & $1$--$6\,\%$\\
  exponential scale $\hat\theta$ & $7.9\,\%$ (median)\\
  first-drop radius & $11$--$13\,\%$\\
  first-drop velocity & up to $50\,\%$\\
  \end{tabular}
  \caption{Dispersion among replicas of the same physical case, seventeen cases
  with two to four realizations each.}
  \label{tab:replicas}
  \end{center}
\end{table*}

The dispersion of $\hat\theta$ within a case, $7.9\,\%$, is to be compared
with the factor $1.25$ (some $25\,\%$) between different cases. That the
first is three times smaller is the quantitative content of the statement that
$\theta$ is a property of the cavity rather than of the realization.

All fits in this paper collapse replicas of a case to a single point before
fitting, their dispersion serving as the error bar of that point; figures show
every realization.

\section{Dependence on the refinement level}
\label{app:levels}

The claims about the census do not need the same resolution, and it is
worth stating which needs what.

The relation $R_{\min}=0.729\Delta$ requires the \emph{widest} available range
in $\Delta$, since its content is the dependence on $\Delta$; all levels
$N=12$ to $15$ are used, which, with a box that grows with $\lambda$, span a
factor $32$ in $\Delta$.

The scale $\theta$ is insensitive to the level: restricting to $N\ge12$,
$N\ge13$ and $N\ge14$ gives $\theta/R_0=0.0390$, $0.0394$ and $0.0418$ on $59$,
$46$ and $21$ cases, and the viscous factor of (\ref{eq:joint}) is the same at
every level. The variation
is consistent with sampling noise and shows no systematic drift.

\begin{figure*}
  \centering
  \includegraphics[width=\textwidth]{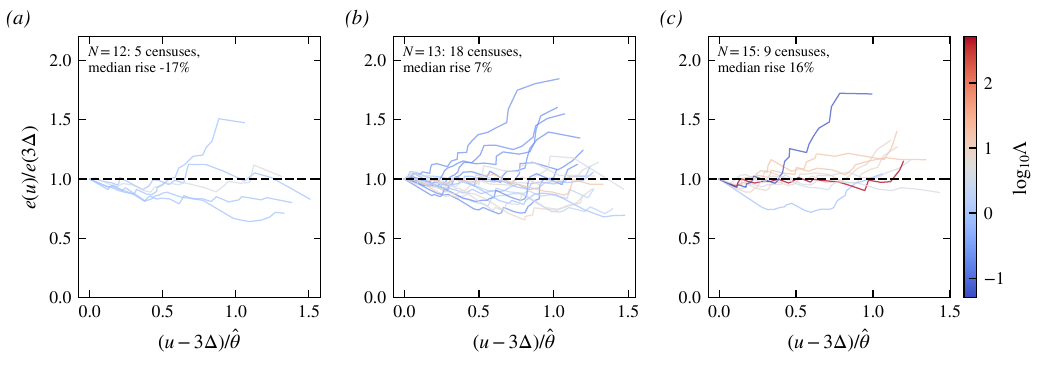}
  \caption{Mean-excess functions divided by their value at the truncation,
  $e(3\Delta)=\hat\theta$, as in figure \ref{fig:exp}(\textit{a}), separated by
  refinement level, (\textit{a}) $N=12$, (\textit{b}) $N=13$, (\textit{c})
  $N=15$, with median rises of $-17$, $7$ and $16\,\%$; colour denotes the confinement $\Lambda$, as in
  figure \ref{fig:exp}. A flat curve is an exponential census. The drift with level is
  visible and is not averaged away: the median rise between the truncation and
  the third quartile grows from $8\,\%$ at $N\ge12$ to $16\,\%$ at $N\ge14$,
  while the number of censuses large enough to be diagnosed
  ($n_{\rm res}\ge40$) falls from $35$ to $12$. Level $N=14$ carries too few
  completed runs for a panel of its own and is quoted in the text only. The
  curves that stand furthest above the horizontal at every level are the
  thin-shell columns $\Lambda<1$, where a second mechanism adds a coarse tail:
  the level dependence and the confinement dependence are distinct effects and
  are seen here to be so.}
  \label{fig:niveles}
\end{figure*}

The shape of the census does drift, and in the direction that flatters the
exponential on coarse grids. The mean-excess function rises by a median $8\,\%$, $11\,\%$ and $16\,\%$
for $N\ge12$, $N\ge13$ and $N\ge14$, on $35$, $30$ and $12$ censuses of at
least forty resolved droplets, and is flat to within $\pm20\,\%$ in $21$, $18$
and $7$ of them (figure \ref{fig:niveles}). A coarse mesh raises
the truncation $3\Delta$, and above a sufficiently high threshold the excesses
of any light-tailed distribution tend to an exponential; coarse meshes
therefore favour apparent memorylessness. Level $N=12$ taken alone goes further
and falls below it, with a median of $-17\,\%$ on five censuses.

The statistics of \S\ref{sec:exp}
are restricted to $N\ge13$ and the drift is quoted rather than averaged away.
It does not converge to zero with the levels available, and the exponential
census is accordingly presented as an indication supported by a consistent
body of evidence, not as a demonstration.

\section{Jet versus film drops in the submicrometre range}
\label{app:jetfilm}

The mechanistic sea-spray generation function of \citet{DeikeReichlPaulot2022} assembles the spray from three modes (jet
drops, and film drops in two sub-processes), each described by a mean number
and a mean radius per bursting bubble, each dressed with a per-event gamma
distribution of order $m_\Gamma$, and the whole integrated over the bubble population.
Their result attributes the submicrometre aerosol almost entirely to film
drops. That attribution is not independent of one unmeasured choice, and this
section makes the dependence explicit.

\subsection{The order of the per-event law decides the outcome}

A gamma density of order $m_\Gamma$, normalized to unit mean, behaves as $x^{m_\Gamma-1}$ as
$x\to0$. The small-size tail of the per-event census is therefore suppressed as
$x^{10}$ for $m_\Gamma=11$, as $x^{3}$ for $m_\Gamma=4$, and not at all for the exponential,
$m_\Gamma=1$. Three values are in circulation for the jet drops of a bursting bubble:
$m_\Gamma=1$, the census measured here (\S\ref{sec:exp}); $m_\Gamma=4$, fitted by \citet{Berny2021} to ensembles of order one hundred realizations; and
$m_\Gamma=11$, adopted by \citet{DeikeReichlPaulot2022} for both
jet and film drops on the grounds that the order ``will only affect the shape at
the tail and peak of the distribution''. It is precisely that part of the curve
which decides the attribution.

Integrating their equations (1)--(5) over the bubble population they use (the
Deane--Stokes double power law, $n_b\propto R_0^{-3/2}$ below the Hinze scale
and $R_0^{-10/3}$ above; \citealt{DeanStok2002}), for seawater at $20\,^\circ$C
and a contaminant Schmidt number $Sc=10^{3}$, the jet-to-film ratio of the
integrated spectrum is given in table \ref{tab:jetfilm}.

\begin{table*}
  \begin{center}
  \def~{\hphantom{0}}
  \begin{tabular}{lcc}
  & \multicolumn{2}{c}{Jet/film, exponential census}\\
  $r_d$ ($\mu$m) & $R_0>60\,\mu$m & $R_0>5\,\mu$m\\[3pt]
  $1.0$  & $0.0004$ & $0.005$\\
  $0.5$  & $0.001$  & $0.031$\\
  $0.2$  & $0.13$   & $6.9$\\
  $0.1$  & $24$     & $1.7\times10^{3}$\\
  $0.05$ & $1.0\times10^{4}$ & $8.6\times10^{5}$\\
  \end{tabular}
  \caption{Ratio of jet to film drops in the integrated spectrum of
  \citet{DeikeReichlPaulot2022}, with their per-event gamma census replaced by
  the exponential census measured here, for two lower limits of the bubble
  population (seawater, $20\,^\circ$C, $Sc=10^3$).}
  \label{tab:jetfilm}
  \end{center}
\end{table*}

With their own $m_\Gamma=11$, by contrast, the same jet mode contributes
$1.8\times10^{-4}$ at $r_d=0.1\,\mu$m against $2.6\times10^{8}$ for the
exponential census, in the same units: twelve orders of magnitude, on the order
of the census alone. With $m_\Gamma=11$ the jet mode does not exist below
a micrometre and the submicrometre aerosol is film by construction.

Two remarks, in opposite directions. The crossover is at $r_d\simeq0.12\,\mu$m
if the bubble population is truncated at $R_0=60\,\mu$m as they truncate it,
and at $\simeq0.23\,\mu$m if the truncation is lowered to $5\,\mu$m, where
the jet mode still operates and the film mode does not, its lower limit being
$100\,\mu$m. And between $0.3$ and $3\,\mu$m the film mode still dominates in
their own parametrization, by two to three orders of magnitude; we do not claim
the jet mode there. That band, however, is where their numbers are weakest:
they declare a factor $2$--$4$ uncertainty on the film modes, the Schmidt
dependence ``remains to be validated experimentally'' in their words, and the
flapping mechanism \citep{JiangEtal2022} is extrapolated down to bubbles of
$100\,\mu$m although the rim of a small bubble need not fragment into ligaments
at all.

\subsection{An independent consistency check}

\begin{figure}
  \centering
  \includegraphics[width=\columnwidth]{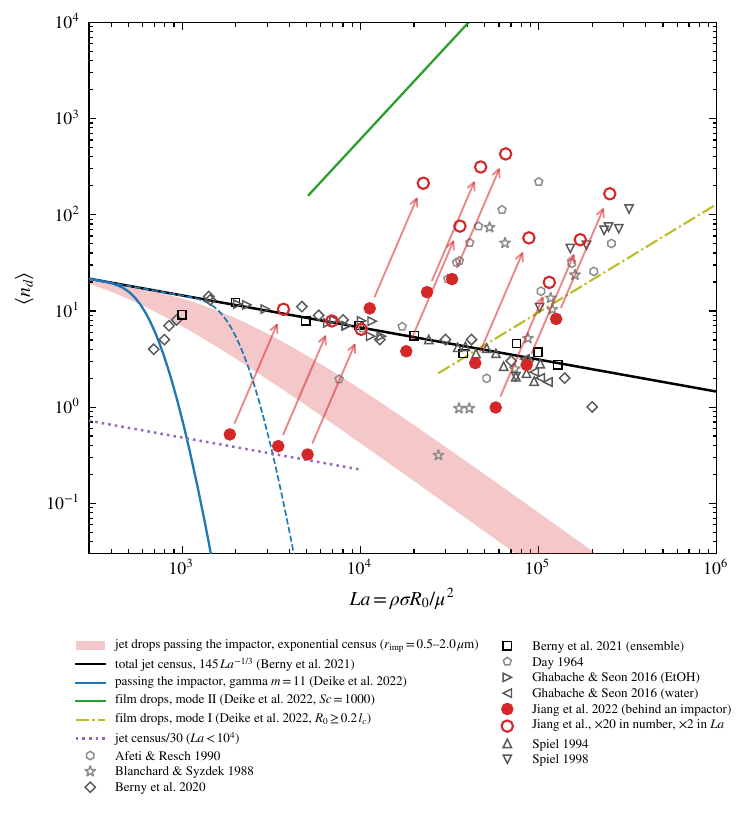}
  \caption{Droplets per bursting bubble: the jet census that passes the
  impactor of \citet{JiangEtal2022}, against what they counted, as a function
  of $La=\rho\sigma R_0/\mu^2$. Grey symbols: published counts
  \citep{Berny2020,Berny2021,Spiel1994,Spiel1998,GhabacheSeon2016,Day1964,
  Afeti1990,BlanchardSyzdek1988}. Filled red: the seawater series of
  \citet{JiangEtal2022}, measured behind an impactor. Open red, with arrows:
  the same points corrected by a factor $20$ in number and $2$ in $La$. Black:
  the total jet census, $145\,La^{-1/3}$ \citep{Berny2021}. Shaded band: the
  fraction of it that passes an impactor of cut $r_{\rm imp}=0.5$--$2\,\mu$m,
  for the exponential census of \S\ref{sec:exp}. Blue: the same for a gamma
  census of order $11$ \citep{DeikeReichlPaulot2022}, solid at
  $r_{\rm imp}=0.5\,\mu$m and dashed at $2\,\mu$m. Green and olive: film drops,
  modes II and I of \citet{DeikeReichlPaulot2022}. Dotted purple: the jet
  census divided by $30$, for $La<10^4$.}
  \label{fig:pnas}
\end{figure}

The point can be tested against a measurement rather than a model. \citet{JiangEtal2022} report submicrometre aerosol from
controlled bubbles and attribute it to film flapping; replying to a
comment~\citep{GananCalvo2022C}, they account for the deficit of their counts
relative to published jet-drop numbers by noting that an impactor placed before
their counter removed the micrometre-sized jet drops, which therefore
``were not measured at all''~\citep{JiangEtal2022R}. That argument is valid only
if the jet population has no appreciable small tail.

With the exponential census, $\theta=cR_0$, the fraction of jet drops below the
impactor cut is $1-e^{-r_{\rm imp}/\theta}$, and for their three smallest
bubbles it is not negligible (figure \ref{fig:pnas} and table \ref{tab:jiang}).

\begin{table*}
  \begin{center}
  \def~{\hphantom{0}}
  \begin{tabular}{ccccc}
  $La$ & $R_0$ ($\mu$m) & $\theta$ ($\mu$m) & Passes & Measured\\[3pt]
  $1852$ & $36.5$ & $1.43$ & $3.49$ & $0.520$\\
  $3462$ & $68.2$ & $2.67$ & $1.64$ & $0.393$\\
  $5059$ & $99.7$ & $3.90$ & $1.02$ & $0.322$\\
  \end{tabular}
  \caption{The three smallest bubbles of \citet{JiangEtal2022}: jet drops per
  bubble that pass their impactor according to the exponential census,
  $\theta=cR_0$, against the aerosol they measured per bubble.}
  \label{tab:jiang}
  \end{center}
\end{table*}

The jet drops that traverse their own impactor exceed, by themselves, the whole
aerosol they counted, by a factor $3$ to $7$. Those three points are also the
only ones in the compilation that depart from the jet law
$\langle n_d\rangle=145\,La^{-1/3}$, by a factor of about twenty, every other
source in the same range lying between $4$ and $14$; and the correction
proposed in the comment (a factor $20$ in number and $2$ in $La$) places
them on that law within $11\,\%$ while overshooting by factors of $17$ to $119$
above $La\simeq2\times10^{4}$, where film flapping is genuinely operative.

The two readings are mutually exclusive and both deny the film attribution
below $La\simeq10^{4}$: either the counts are low, or what was counted there
was jet. Discriminating them requires counting the submicrometre aerosol of
bubbles with $La=10^{3}$--$10^{4}$ without an impactor, which is the experiment
\S\ref{sec:discussion} calls for.

\section{The campaign}
\label{app:campaign}

The figures quoted in the body refer to the confined family, with the
resolved census defined by $R_d\ge3\Delta$ and replicas of a physical case
collapsed to a single point before any fit; the runs that enter each result
are listed below.

\subsection{Inventory}

The campaign comprises $285$ realizations at levels $N=12$ to $15$, every one
of them kept: $222$ at single tangency and $63$ at double tangency. Of the
single-tangency runs, $155$ terminated normally, $29$ were stopped, $16$
aborted, $8$ diverged and $14$ are incomplete or have no log; of the
double-tangency runs, $48$ terminated normally, $7$ were stopped, $6$ aborted
and $2$ diverged. A run that did not terminate normally still counts wherever
it reached the stage a result needs. Each result therefore rests on its own
subset, listed in table \ref{tab:subsets}.

\begin{table*}
  \begin{center}
  \begin{tabular}{llc}
  Result & Criterion & Runs (cases)\\[3pt]
  Ejection boundary (\S\ref{sec:bound}) & escape, or no escape by $t=1.5$ & $190$\\
  \quad shown in figure \ref{fig:map} & within $0.025$ of (\ref{eq:law}) & $73$\\
  Clock and energy (\S\S\ref{sec:clock}--\ref{sec:energy}) & kinetic-energy maximum, all columns & ($119$)\\
  Ejected mass (\S\ref{sec:mass}) & completed ejection, $\delta\ge0.1$, $\Lambda\ge1/4$ & $74$ ($71$)\\
  First droplet (\S\ref{sec:crest}) & escape, $N=13$ or $15$, $\Lambda\ge1$ & $72$ ($48$)\\
  Census (\S\ref{sec:census}) & $t_{\rm fin}\ge1.2$, radii recorded & $136$ ($113$)\\
  \quad mesh floor (figure \ref{fig:mesh}) & at least ten resolved droplets & $106$ ($83$)\\
  \quad exponential census (figure \ref{fig:exp}) & $N\ge13$, $\ge20$ or $\ge40$ resolved & $59$ / $30$\\
  \quad scale and volume against $\delta$ & $\delta\ge0.1$, $\Lambda\ge1/4$, $\ge10$ resolved & $101$ ($78$)\\
  Double tangency (\S\ref{sec:twomouths}) & completed & $48$\\
  \end{tabular}
  \caption{The subset of the campaign on which each result rests.}
  \label{tab:subsets}
  \end{center}
\end{table*}

\subsection{The mesh floor}

Over the $83$ cases with at least ten resolved droplets, the smallest recorded
fragment obeys $R_{\min}=0.729\,\Delta$ with an interquartile range of
$2.3\,\%$ of the median, a coefficient of variation of $3.3\,\%$, and not one
case outside $[0.5,1.5]\,\Delta$, across a factor $32$ in $\Delta$. Normalized
by $\ell_\mu$ instead of by $\Delta$, the same quantity spans three and a half
decades: it follows the mesh and not the fluid.

\bibliographystyle{apsrev4-2}
\bibliography{abc}

\end{document}